\documentclass[a4paper,12pt]{article}
\usepackage{amsmath}
\usepackage{amsfonts}
\usepackage{amssymb}
\usepackage{latexsym}
\usepackage{epsfig}
\usepackage{graphicx}
\usepackage{oldgerm}
\usepackage{theorem}

\def\Z{\mathbb{Z}}
\def\N{\mathbb{N}}
\def\C{\mathbb{C}}
\def\R{\mathbb{R}}

\def\P{\mathbb{P}}

\def\cP{{\cal P}}

\theorembodyfont{\itshape}

\newtheorem{thm}{Theorem}[section]
\newtheorem{lem}[thm]{Lemma}
\newtheorem{cor}[thm]{Corollary}
\newtheorem{prop}[thm]{Proposition}

\newcommand{\SSC}[1]{\section{#1}\setcounter{equation}{0}}
\newcommand{\qed}{\hbox{\rule[-2pt]{3pt}{6pt}}}

\begin{document}

\title{\bf 
Excursion Processes
Associated with \\
Elliptic Combinatorics
}
\author{
Hiroya Baba
\footnote{
Department of Physics,
Faculty of Science and Engineering,
Chuo University, 
Kasuga, Bunkyo-ku, Tokyo 112-8551, Japan;
e-mail: baba@phys.chuo-u.ac.jp}, 
Makoto Katori
\footnote{
Department of Physics,
Faculty of Science and Engineering,
Chuo University, 
Kasuga, Bunkyo-ku, Tokyo 112-8551, Japan;
e-mail: katori@phys.chuo-u.ac.jp} 
}
\date{17 April 2018}
\pagestyle{plain}
\maketitle

\begin{abstract}
Researching elliptic analogues for equalities and formulas
is a new trend in enumerative combinatorics
which has followed the previous trend of studying
$q$-analogues.
Recently Schlosser proposed a lattice path model
in the square lattice with a family of 
totally elliptic weight-functions 
including several complex parameters
and discussed an elliptic extension of the binomial theorem.
In the present paper, we introduce a family of
discrete-time excursion processes on $\Z$ 
starting from the origin and returning to the origin
in a given time duration $2T$
associated with Schlosser's elliptic combinatorics.
The processes are inhomogeneous both in space and time
and hence expected to provide new models
in non-equilibrium statistical mechanics.
By numerical calculation we show that
the maximum likelihood trajectories
on the spatio-temporal plane
of the elliptic excursion processes
and of their reduced trigonometric versions
are not straight lines in general
but are nontrivially curved depending on parameters.
We analyze asymptotic probability laws 
in the long-term limit $T \to \infty$
for a simplified trigonometric version 
of excursion process.
Emergence of nontrivial curves of trajectories
in a large scale of space and time
from the elementary elliptic weight-functions exhibits
a new aspect of elliptic combinatorics.

\vskip 0.2cm



\end{abstract}
\vspace{3mm}

\SSC
{Introduction} \label{sec:introduction}

Recently {\it elliptic extensions} of special functions
and combinatorial identities have been extensively studied
\cite{War02,Spi02,KN03,GR04,SIGMA_special}.
In the present paper, we will report one of the trials
to import such a trend in 
enumerative combinatorics and integrable systems
\cite{DJKMO87,FT97}
into the study of stochastic processes describing
non-equilibrium statistical mechanics models \cite{Kat_Springer}.
In the previous papers \cite{Kat15,Kat16,Kat17},
one of the present authors studied the
elliptic extensions of interacting particle systems
in continuous space-time
called the elliptic Dyson models, 
using the elliptic determinantal identities
given by Rosengren and Schlosser \cite{RS06}. 
Here we consider more fundamental models
defined on a discrete spatio-temporal plane;
excursion processes of a single particle
associated with elliptic combinatorics \cite{Sch07,Sch11}.

As mentioned in Section 5.11 in \cite{Kra05},
the currently spreading ``elliptic disease" has followed
the previous ``$q$-disease" (see also Footnote 20 in \cite{Kra05}).
When we are infected by the ``$q$-disease", 
we replace every positive integer $n$ by the {\it $q$-number}
\[
[n]_q \equiv 1+q+q^2+ \dots + q^{n-1} = \frac{1-q^n}{1-q},
\]
and shifted factorial 
$(a)_k=a(a+1) \dots (a+k-1)$ 
by {\it $q$-shifted factorials}
\begin{equation}
(\alpha; q)_k = (1-\alpha)(1-\alpha q) \dots (1-\alpha q^{k-1})
\quad \mbox{with $\alpha=q^a$}.
\label{eqn:q_shift}
\end{equation}
By these replacements in a classical identity we will
hopefully obtain a new identity, which is regarded
as the $q$-analogue of the original identity
\cite{Kra99,GR04,Kra05,Cha16}.
Now, if we are infected by the ``elliptic disease",
we would replace every term in the form $1-\alpha q^{\ell}$
in (\ref{eqn:q_shift}) by its elliptic analogue
\begin{equation}
\theta(\alpha q^{\ell}; p)
= \prod_{j=0}^{\infty} (1-p^j \alpha q^{\ell})
(1-p^{j+1}/(\alpha q^{\ell})),
\quad \ell =0,1,2, \dots, 
\label{eqn:theta0}
\end{equation}
where $p \in \C$ is a fixed complex number
with $|p| < 1$. 
By definition, if $p=0$,
$\theta(\alpha q^{\ell}; p)$ reduces to $1- \alpha q^{\ell}$.

In \cite{Sch07} Schlosser introduced a lattice path model
in $\Z^2$ consisting of positive directed unit
vertical and horizontal steps with an elliptic weight function.
Here we express his model
as a lattice path model on a spatio-temporal plane,
since we would like to discuss stochastic processes
in this paper. 
Instead of the function $\theta$ given by (\ref{eqn:theta0}),
we use the {\it Jacobi theta function} $\vartheta_1(v; \tau)$
which is obtained from $\theta$ by multiplying a proper factor as
\begin{equation}
\vartheta_1(v; \tau)=
i e^{\pi i (\tau/4- v)} \prod_{n=1}^{\infty} (1-e^{2 \pi i n \tau})
\theta(e^{2 \pi i v}; e^{2 \pi i \tau}),
\label{eqn:varthata_1}
\end{equation}
$i=\sqrt{-1}$, $v, \tau \in \C$ with $\Im \tau > 0$.
It is holomorphic for $|v| < \infty$ and
has an expansion formula
\[
\vartheta_1(v; \tau)
= 2 \sum_{n=1}^{\infty} (-1)^{n-1}
e^{\pi i \tau(n-1/2)^2} \sin \{(2n-1) \pi v \}.
\]
From this expression, we can see that 
$\vartheta_1(v; \tau) \in \R$, if
$v \in \R$ and $\tau \in i \R$ with $\Im \tau > 0$, and
\begin{equation}
\vartheta_1(v; \tau) \sim 2 e^{\pi i \tau/4} \sin (\pi v), 
\quad \mbox{when $\Im \tau \to +\infty$}.
\label{eqn:limit_1}
\end{equation}
(Note that the present functions 
$\vartheta_1(v; \tau)$ is denoted by
$\vartheta_1(\pi v, e^{\pi i \tau})$ in \cite{WW27}.)
Schlosser considered an ensemble of lattice paths
on a spatio-temporal plane
\begin{equation}
\Lambda=\{(t, x) : t \in \N_0, x \in \Z, t+x \in 2 \Z\},
\label{eqn:Lambda}
\end{equation}
where $\N=\{1,2, \dots\}$ and $\N_0=\N \cup \{0\}$.
The elementary steps consist of 
a rightward step
$(t-1, x-1) \to (t, x)$ with weight $q(t,x)$ and
a leftward step
$(t-1, x+1) \to (t, x)$ with weight 1, 
where
\begin{align}
q(t,x) ={}& q(t, x; r, \alpha, \beta, \kappa)
\nonumber\\
={}& \frac{\vartheta_1 \left( \frac{2\alpha+3t-x}{2 \pi r} ; i \kappa \right)}
{\vartheta_1 \left( \frac{2\alpha+t+x}{2 \pi r} ; i \kappa \right)}
\frac{\vartheta_1 \left( \frac{\beta+t+x}{\pi r} ; i \kappa \right) 
\vartheta_1 \left( \frac{\beta+t+x-1}{\pi r} ; i \kappa \right)}
{\vartheta_1 \left( \frac{2\beta+3t+x}{2 \pi r} ; i \kappa \right) 
\vartheta_1 \left( \frac{2 \beta+3t+x-2}{2 \pi r} ; i \kappa \right)}
\nonumber\\
&\times 
\frac{\vartheta_1 \left( \frac{2(\alpha-\beta)-t-x+2}{2 \pi r} ; i \kappa \right) 
\vartheta_1 \left( \frac{2(\alpha-\beta)-t-x}{2 \pi r} ; i \kappa \right)}
{\vartheta_1 \left( \frac{\alpha-\beta-x+1}{\pi r}; i \kappa \right)
\vartheta_1 \left( \frac{\alpha-\beta-x}{\pi r} ; i \kappa \right)},
\label{eqn:q_1}
\end{align}
which depends on four parameters,
$\alpha, \beta \in \C$, $r, \kappa >0$.
For two spatio-temporal points $(s, x), (t, y) \in \Lambda$,
if $s \leq t$ and $|y-x| \leq t-s$, 
a {\it lattice path} $\varpi$ is defined as a sequence of 
spatio-temporal points 
$(u, z_u) \in \Lambda, u=s, s+1, \dots, t$ 
such that 
$z_u-z_{u-1} \in \{-1, 1\}$, $u=s+1, s+2, \dots, t$ with
$z_s=x$ and $z_t=y$.
To each lattice path the following weight is assigned,
\[
q(\varpi)= \prod_{u=s+1}^t 
\{{\bf 1}(z_u-z_{u-1} =1) q(u, z_u) + {\bf 1}(z_u-z_{u-1}=-1)\}
=\prod_{\substack{u=s+1, \dots, t, \cr z_u-z_{u-1}=1}} q(u, z_u),
\]
where ${\bf 1}(\omega)$ is an indicator function of a condition $\omega$;
${\bf 1}(\omega)=1$ if $\omega$ is satisfied and ${\bf 1}(\omega)=0$
otherwise.
Let $\Pi((s, x) \to (t, y))$ be the collection of all lattice paths
starting from $(s, x)$ and arriving at $(t, y)$ on $\Lambda$.
Then the {\it transition weight} from $(s, x)$ to $(t, y)$ is defined by 
the sum
\begin{equation}
Q(s, x; t, y) = \sum_{\varpi \in \Pi((s, x) \to (t, y))} q(\varpi).
\label{eqn:Q_1}
\end{equation}
If $\Pi((s, x) \to (t, y)) = \emptyset$, then  we put
$Q(s, x; t, y)=0$. 
Schlosser made a special choice of weight for 
an elementary rightward step as (\ref{eqn:q_1})
so that the transition weight (\ref{eqn:Q_1}) can be factorized as
\begin{align}
Q(s, x; t, y)
={}& Q(s, x; t, y; r, \alpha, \beta, \kappa)
\nonumber\\
={}& \frac{
\prod_{u=1}^{t-s} \vartheta_1 \left( \frac{u}{\pi r}; i \kappa \right)}
{\prod_{u=1}^{\{(t-s)+(y-x)\}/2} \vartheta_1 \left( \frac{u}{\pi r} ; i \kappa \right)
\prod_{u=1}^{\{(t-s)-(y-x)\}/2} \vartheta_1 \left( \frac{u}{\pi r} ; i \kappa \right)}
\nonumber\\
&\times
\prod_{u=(s+x)/2+1}^{(t+y)/2}
\frac{\vartheta_1 \left( \frac{\alpha+\{(t-y)+(s-x)\}/2+u}{\pi r}; i \kappa \right)}
{\vartheta_1 \left( \frac{\alpha+u}{\pi r}; i \kappa \right)}
\nonumber\\
&\times
\frac{\prod_{u=(s-x)/2+1}^{(t-y)/2}
\vartheta_1 \left( \frac{\beta+\{(t+y)+(s+x)\}/2+u}{\pi r}; i \kappa \right) 
\prod_{u=(s+x)+1}^{t+y}
\vartheta_1 \left( \frac{\beta+u}{\pi r}; i \kappa \right)}
{\prod_{u=(3s+x)/2+1}^{(3t+y)/2}
\vartheta_1 \left( \frac{\beta+u}{\pi r}; i \kappa \right)}
\nonumber\\
&\times
\prod_{u=-(t+y)/2+1}^{-(s+x)/2}
\frac{ \vartheta_1 \left( \frac{\alpha-\beta+u}{\pi r}; i \kappa \right) 
\vartheta_1 \left( \frac{\alpha-\beta-1+u}{\pi r}; i \kappa \right)}
{\vartheta_1 \left( \frac{\alpha-\beta+(t-y)/2+u}{\pi r}; i \kappa \right) 
\vartheta_1 \left( \frac{\alpha-\beta+(s-x-2)/2+u}{\pi r}; i \kappa \right)}.
\label{eqn:Q_2}
\end{align}
This fact can be proved by showing that (\ref{eqn:Q_2}) solves
the recursion relation 
\[
Q(s,x; t, y) = Q(s, x; t-1, y-1) q(t, y)+Q(s, x; t-1, y+1),
\quad s \leq t, |y-x| \leq t-s,
\]
with (\ref{eqn:q_1}), 
and it was verified using the {\it addition formula of theta functions}
(see Example 5 on page 451 of \cite{WW27}),
\begin{align}
& \vartheta_1(x+y; \tau) \vartheta_1(x-y; \tau) \vartheta_1(u+v; \tau) \vartheta_1(u-v;\tau)
\nonumber\\
& \quad
- \vartheta_1(x+v; \tau) \vartheta_1(x-v; \tau) \vartheta_1(y+u; \tau) \vartheta_1(-y+u;\tau)
\nonumber\\
& \qquad 
= \vartheta_1(y+v; \tau) \vartheta_1(y-v; \tau) \vartheta_1(x+u; \tau) \vartheta_1(x-u;\tau).
\label{eqn:Riemann1}
\end{align}
See \cite{Koo14} for more details of (\ref{eqn:Riemann1}).
Concerning the elementary weight (\ref{eqn:q_1}) 
and the transition weight (\ref{eqn:Q_1}), 
we note the following three points.
\begin{description}
\item{(i)} \quad
As a complex function of $t, x, \alpha, \beta$,
the elementary weight (\ref{eqn:q_1}) is 
{\it totally elliptic} \cite{Spi02} in the sense that
all free parameters $t, x, \alpha$, and $\beta$ 
viewed as complex variables
have equal periods of doubly periodicity.

\item{(ii)} \quad
By definition of lattice path models,
the transition weight (\ref{eqn:Q_1}) satisfies the
{\it Chapman-Kolmogorov equation}
\begin{equation}
Q(s, x; t, y)= \sum_{z \in \Z}
Q(s, x; u, z) Q(u, z; t, y)
\label{eqn:CK1}
\end{equation}
for $0 < s < u <t$, $x, y \in \Z$.
Schlosser \cite{Sch07} proved that
if we use the expression (\ref{eqn:Q_2}),
a variant of Frenkel and Turaev's ${_{10}}V_9$ summation
formula of elliptic hypergeometric functions \cite{FT97}
is derived from (\ref{eqn:CK1}),
which is the elliptic extension of
Jackson's very-well-poised balanced
${_8}\phi_7$ summation formula
for $q$-hypergeometric functions \cite{GR04}.

\item{(iii)} \quad
For (\ref{eqn:limit_1}), 
if we take the limit $\kappa \to \infty$, 
the elliptic weight-functions (\ref{eqn:q_1}) and (\ref{eqn:Q_2}) become
trigonometric weight-functions.
Then if we put $\alpha=i \widehat{\alpha}$,
$\beta=i \widehat{\beta}$ with $\widehat{\alpha}, \widehat{\beta} \in \R$,
and take the limit $\widehat{\beta} \to \infty$ and then
$\widehat{\alpha} \to - \infty$
(or $\widehat{\alpha} \to \infty$, $\widehat{\beta} \to \infty$
in this order), we will have the following,
\begin{align}
q(t,x) &\to q^{(t-x)/2}, 
\nonumber\\
Q(s,x; t,y) &\to
\left[ \begin{array}{c} t-s \cr \{(t-s)+(y-x)\}/2 \end{array} \right]_q
q^{\{(t-s)+(y-x)\}(s-x)/4}
\label{eqn:Q_limit1}
\end{align}
with $q=e^{2 i/r}$, where the {\it $q$-binomial coefficient} is defined by
\[
\left[ \begin{array}{c} n \cr k \end{array} \right]_q
=\frac{(q; q)_n}{(q; q)_k  (q; q)_{n-k}}.
\]
Moreover, if we take the further limit $r \to \infty$, 
then $q \to 1$, $q^{(t-s)/2} \to 1$ and 
(\ref{eqn:Q_limit1}) is reduced to the usual 
binomial coefficient,
\begin{equation}
Q(s,x;t,y) \to
\binom{t-s}{\{(t-s)+(y-x)\}/2}.
\label{eqn:Q_limit2}
\end{equation}
\end{description}
The above three points verify the fact that
Schlosser's elliptic lattice-path model 
provides a standard elliptic extension
of the binomial theorem and thus
it will give a basis to build a theory
of elliptic combinatorics \cite{Sch07,Sch11,SY17}.

From the viewpoint of probability theory and statistical mechanics, however, 
we find some difficulties in Schlosser's elliptic 
lattice path model and the elliptic binomial theorem.
The choice of the weight (\ref{eqn:q_1}) for the elementary rightward 
step is very suitable in order to apply the addition formula of theta functions
(\ref{eqn:Riemann1}) and thus 
the transition weight (\ref{eqn:Q_1}) can be
perfectly factorized as (\ref{eqn:Q_2}).
On the other hand, we cannot expect 
any compact formula for the sum
\[
c(s, x; t) = \sum_{y \in \Z} Q(s, x; t, y)
\]
for $(s, x) \in \Lambda$, $s < t \in \N$.
Since the transition weight (\ref{eqn:Q_2})
is inhomogeneous both in space and time
contrary to its classical limit (\ref{eqn:Q_limit1}),
the sum $c(s, x; t)$ indeed depends on $s, x$, and $t$.
Therefore, if we simply define the transition probability
as $p(s, x; t, y)=Q(s, x; t, y)/c(s, x; t)$,
which is well normalized as
$\sum_{y \in \Z} p(s, x; t, y)=1$,
the Chapman-Kolmogorov equation
will not hold for $p(s,x;t,y)$. 
In addition to this normalization problem,
we have to clarify the positivity condition
to define proper transition probability 
from the transition weight (\ref{eqn:Q_2})
in order to construct a stochastic process.
In the present paper, we introduce a family of discrete-time
{\it excursion processes} in $\Z$ associated with Schlosser's
totally elliptic weight functions (\ref{eqn:q_1}) and (\ref{eqn:Q_2})
by solving the above mentioned problems.
We consider the excursion processes 
starting from the origin and returning to the origin
in a given time duration $2T$, $T \in \N$.
The processes are inhomogeneous both in space and time
and hence expected to provide new models
in non-equilibrium statistical mechanics.
By numerical calculation we show that
the {\it maximum likelihood trajectories} of the elliptic excursion processes
on the spatio-temporal plane are not straight lines in general
and are curved depending on parameters
(see Figs. \ref{fig:simp_tri_T=100} and \ref{fig:tri_T=100}). 
We will analyze asymptotic probability laws 
in the long-term limit $T \to \infty$
for a reduced model which we call 
a simplified trigonometric excursion process.
It should be noted that the elementary weight $q(t,x)$
is a simple function of $(t,x)$ as shown in Figs. \ref{fig:simp_tri_q} and
\ref{fig:tri_q}.
Emergence of nontrivial curves on the spatio-temporal plane
is highly nontrivial and has not been expected
in the previous study.

The paper is organized as follows.
In Section \ref{sec:preliminaries}, 
first we explain how to derive the spatio-temporal
expressions for weights (\ref{eqn:q_1}) and (\ref{eqn:Q_2})
from Schlosser's formulas originally given for 
the lattice path models in $\Z^2$.
The excursion processes are defined for finite time
durations $2T, T \in \N$.
In Section \ref{sec:parameterization}, we adopt
a $T$-depending parameterization
for $\alpha$ and $\beta$ to make the expressions
of measures for trajectories be simpler,
but this parameterization is not yet essential,
since still it contains arbitrary parameters
$\alpha_0$ and $\beta_0$.
In our expressions using the Jacobi theta functions
and trigonometric functions, it is obvious that
if $\alpha_0, \beta_0 \in \R$, $r, \kappa >0$,
the obtained measures for trajectories are real-valued, but they are
not non-negative definite; they are {\it signed measures} 
in general.
In Section \ref{sec:reduction},
we show the reductions to trigonometric and classical measures for trajectories 
from the elliptic measures by taking proper
limits of parameters $r, \alpha_0, \beta_0$ and $\kappa$.
We define two levels of trigonometric processes;
the trigonometric excursion processes
and its simplified version.
Section \ref{sec:positivity} is devoted to giving
sufficient conditions to make
the measures for trajectories 
be non-negative
for the elliptic excursion processes 
(Theorem \ref{thm:positivity_main})
and for two kinds of trigonometric excursion processes
(Corollaries \ref{thm:positivity_tri} and
\ref{thm:positivity_simp_tri}).
Then the probability measures
$\P^{0,0}_{2T}$ for the excursion processes
$X(t), t \in \{0,1,\dots, 2T\}$ are well defined.
In Section \ref{sec:trajectories},
results of numerical calculation are reported for the
single-time probability distributions
$\P^{0,0}_{2T}(X(t)=t)$ for each time $t \in \{0,1,\dots, 2T\}$.
We show that the maximum likelihood trajectories
$x=x_{2T}^{\rm max}(t), t \in \{0,1, \dots, 2T\}$
on the spatio-temporal plane
exhibit nontrivial curves depending on
the parameters and the level of reductions.
We studied the processes 
by increasing the time duration $T$.
The numerical results suggest the existence
of {\it scaled limit trajectories} defined by
\begin{equation}
\lim_{T \to \infty} 
\left\{ \frac{x_{2T}^{\rm max}(sT)}{T} :
0 \leq s \leq 2 \right\}
\label{eqn:limit_traj}
\end{equation}
for our excursion processes.
In Section \ref{sec:asymptotic},
we concentrate on the simplified
trigonometric excursion processes
and analyze the asymptotic probability laws
in the long-time limit $T \to \infty$.
We prove that the {\it large deviation principle} 
is established for $T \to \infty$
by giving an integral representation
for the {\it rate function} (Lemma \ref{thm:rate_func}),
and characterize the scaled limit trajectories (\ref{eqn:limit_traj})
(Theorem \ref{thm:trajectory}).
At the time $t=T$, the {\it central limit theorem} 
for the fluctuation of trajectory is also
proved (Corollary \ref{thm:CLT}).
The proofs of theorems are given in Appendices
A and B.
Concluding remarks are given in Section \ref{sec:concluding}. 

\SSC
{Preliminaries} \label{sec:preliminaries}
\subsection{Rewriting of Schlosser's results} 
\label{sec:rewriting}

First we explain how to obtain the formulas
(\ref{eqn:q_1}) and (\ref{eqn:Q_2}) from the results 
reported in \cite{Sch07}.
Schlosser introduced a lattice path model in $\Z^2$.
The elliptic weight function on horizontal edges
$(n-1, m) \to (n, m)$ of $\Z^2$ is given by
Eq. (2.2) in \cite{Sch07},
which was expressed by the
`modified theta function' (\ref{eqn:theta0}) 
including arbitrary complex parameters
$a, b, q, p$ with $q \not=0$ and $|p| < 1$.
(The weight on vertical edges is fixed to be 1.)
We put
\begin{equation}
a=e^{2 i \alpha/r}, \quad
b=e^{2 i \beta/r}, \quad
q=e^{2 i /r}, \quad
p=e^{2 \pi i \tau}=e^{-2 \pi \kappa}
\label{eqn:para1}
\end{equation}
with $r >0$ and $\Im \tau \equiv \kappa >0$ and use the relation
(\ref{eqn:varthata_1}).
Then the weight function on horizontal edges $(n-1, m) \to (n, m)$
of $\Z^2$ is rewritten using the Jacobi theta functions as
\begin{align}
w(n,m)  
={}&
\frac{
\vartheta_1 \left( \frac{\alpha+n+2m}{\pi r}; i \kappa \right)
\vartheta_1 \Big( \frac{\beta+2n}{\pi r}; i \kappa \Big)
\vartheta_1 \left( \frac{\beta+2n-1}{\pi r}; i \kappa \right)
}
{
\vartheta_1 \left( \frac{\alpha+n}{\pi r}; i \kappa \right)
\vartheta_1 \left( \frac{\beta+2n+m}{\pi r}; i \kappa \right)
\vartheta_1 \left( \frac{\beta+2n+m-1}{\pi r}; i \kappa \right)
}
\nonumber\\
&
\times
\frac{ 
\vartheta_1 \left( \frac{\alpha-\beta+1-n}{\pi r}; i \kappa \right)
\vartheta_1 \left( \frac{\alpha-\beta-n}{\pi r}; i \kappa \right)
}
{
\vartheta_1 \left( \frac{\alpha-\beta+1+m-n}{\pi r}; i \kappa \right)
\vartheta_1 \left( \frac{\alpha-\beta+m-n}{\pi r}; i \kappa \right)
}.
\label{eqn:w_1}
\end{align}
Let $w(\cP((\ell, k) \to (n,m)))$ be the generating function of
paths running from $(\ell, k) \in \Z^2$ to
$(n,m) \in \Z^2$ in Schlosser's lattice path model.
The following is just a rewriting of Theorem 2.1
of Schlosser \cite{Sch07} using the Jacobi theta function (\ref{eqn:varthata_1}) 
with the parameterization (\ref{eqn:para1}).
\begin{thm}
[Schlosser \cite{Sch07}]
\label{thm:Schlosser_2_1}
The recursion relation of the generating function
of the lattice paths
\begin{align}
&w(\cP((\ell,k)\to(n,m)))
\nonumber\\
&=w(\cP((\ell,k)\to(n,m-1)))
+w(\cP((\ell,k)\to(n-1,m))) w(n,m)
\label{eqn:recursion_elliptic}
\end{align}
with the weight function (\ref{eqn:w_1}) is solved by
\begin{align}
&w(\cP((\ell,k)\to(n,m)))
\nonumber\\
&\quad = 
\frac{
\prod_{u=1}^{n-\ell+m-k} 
\vartheta_1 \left( \frac{u}{\pi r}; i \kappa \right)
} 
{
\prod_{u=1}^{n-\ell} 
\vartheta_1 \left( \frac{u}{\pi r}; i \kappa \right)
\prod_{u=1}^{m-k} 
\vartheta_1 \left( \frac{u}{\pi r}; i \kappa \right)
}
\prod_{u=\ell+1}^n 
\frac{ 
\vartheta_1 \left( \frac{\alpha+m+k+u}{\pi r}; i \kappa \right)
}
{
\vartheta_1 \left( \frac{\alpha+u}{\pi r}; i \kappa \right)
}
\nonumber\\
&\qquad \times
\frac{ \prod_{u=k+1}^m 
\vartheta_1 \left( \frac{\beta+n+\ell+u}{\pi r}; i \kappa \right)
\prod_{u=2 \ell+1}^{2n} \vartheta_1 \left( \frac{\beta+u}{\pi r}; i \kappa \right)
}
{\prod_{u=2\ell+k+1}^{2n+m} 
\vartheta_1 \left( \frac{\beta+u}{\pi r}; i \kappa \right)
}
\nonumber\\
&\qquad \times 
\prod_{u=-n+1}^{-\ell}
\frac{\vartheta_1 \left( \frac{\alpha-\beta+u}{\pi r}; i \kappa \right)
\vartheta_1 \left( \frac{\alpha-\beta-1+u}{\pi r}; i \kappa \right)}
{\vartheta_1 \left( \frac{\alpha-\beta+m+u}{\pi r}; i \kappa \right)
\vartheta_1 \left( \frac{\alpha-\beta+k-1+u}{\pi r}; i \kappa \right)}.
\label{eqn:wP_1}
\end{align}
\end{thm}
\noindent{\it Proof.} \,
Insert (\ref{eqn:w_1}) and (\ref{eqn:wP_1}) into (\ref{eqn:recursion_elliptic}).
Note that
\begin{align*}
& \prod_{u=\ell+1}^n \vartheta_1 \left( \frac{\alpha+m-1+k+u}{\pi r}; i \kappa \right)
=\prod_{u=\ell}^{n-1} \vartheta_1\left( \frac{\alpha+m+k+u}{\pi r}; i \kappa \right),
\nonumber\\
& \prod_{u=-n+1}^{-\ell} \vartheta_1 \left( \frac{\alpha-\beta+m-1+u}{\pi r}; i \kappa \right)
= \prod_{u=-n}^{-\ell-1} \vartheta_1 \left( \frac{\alpha-\beta+m+u}{\pi r}; i \kappa \right),
\nonumber\\
& \prod_{u=k+1}^m \vartheta_1 \left( \frac{\beta+n-1+\ell+u}{\pi r}; i \kappa \right)
=\prod_{u=k}^{m-1} \vartheta_1 \left( \frac{\beta+n+\ell+u}{\pi r}; i \kappa \right),
\end{align*}
and divide both sides of the equation
by the common factor
\begin{align*}
& \frac{
\prod_{u=1}^{n-\ell+m-k-1} 
\vartheta_1 \left( \frac{u}{\pi r}; i \kappa \right)
} 
{
\prod_{u=1}^{n-\ell-1} 
\vartheta_1 \left( \frac{u}{\pi r}; i \kappa \right)
\prod_{u=1}^{m-k-1} 
\vartheta_1 \left( \frac{u}{\pi r}; i \kappa \right)
}
\prod_{u=\ell+1}^{n}
\frac{ 
\vartheta_1 \left( \frac{\alpha+m+k+u}{\pi r}; i \kappa \right)
}
{
\vartheta_1 \left( \frac{\alpha+u}{\pi r}; i \kappa \right)
}
\nonumber\\
&\times
\frac{\prod_{u=k+1}^{m-1}
\vartheta_1 \left( \frac{\beta+n+\ell+u}{\pi r}; i \kappa \right)
\prod_{u=2 \ell+1}^{2n} \vartheta_1 \left( \frac{\beta+u}{\pi r}; i \kappa \right)
}
{\prod_{u=2\ell+k+1}^{2n+m-1} 
\vartheta_1 \left( \frac{\beta+u}{\pi r}; i \kappa \right)
}
\nonumber\\
&\times \prod_{u=-n+1}^{-\ell}
\frac{\vartheta_1 \left( \frac{\alpha-\beta+u}{\pi r}; i \kappa \right)
\vartheta_1 \left( \frac{\alpha-\beta-1+u}{\pi r}; i \kappa \right)}
{\vartheta_1 \left( \frac{\alpha-\beta+m+u}{\pi r}; i \kappa \right)
\vartheta_1 \left( \frac{\alpha-\beta+k-1+u}{\pi r}; i \kappa \right)}.
\end{align*}
Then we obtain the equation
\begin{align*}
&\frac{\vartheta_1 \left(\frac{n-\ell+m-k}{\pi r}; i \kappa \right) 
\vartheta_1 \left( \frac{\beta+n+\ell+m}{\pi r}; i \kappa \right)}
{\vartheta_1 \left( \frac{n-\ell}{\pi r}; i \kappa \right) 
\vartheta_1 \left( \frac{m-k}{\pi r}; i \kappa \right)
\vartheta_1 \left( \frac{\beta+2n+m}{\pi r}; i \kappa \right)}
\nonumber\\
&= 
\frac{\vartheta_1 \left(\frac{\alpha+m+k+\ell}{\pi r}; i \kappa \right)
\vartheta_1 \left(\frac{\alpha-\beta+m-\ell}{\pi r}; i \kappa \right)}
{\vartheta_1 \left(\frac{n-\ell}{\pi r}; i \kappa \right)
\vartheta_1 \left( \frac{\alpha+m+k+n}{\pi r}; i \kappa \right)
\vartheta_1 \left(\frac{\alpha-\beta+m-n}{\pi r}; i \kappa \right)}
\nonumber\\
&+ 
\frac{\vartheta_1 \left(\frac{\alpha+n+2m}{\pi r}; i \kappa \right)
\vartheta_1 \left(\frac{\beta+n+\ell+k}{\pi r}; i \kappa \right)
\vartheta_1 \left(\frac{\alpha-\beta+k-n}{\pi r}; i \kappa \right)}
{\vartheta_1 \left(\frac{m-k}{\pi r}; i \kappa \right)
\vartheta_1 \left( \frac{\alpha+m+k+n}{\pi r}; i \kappa \right)
\vartheta_1 \left(\frac{\beta+2n+m}{\pi r}; i \kappa \right)
\vartheta_1 \left(\frac{\alpha-\beta+m-n}{\pi r}; i \kappa \right)}.
\end{align*}
If we multiply both sides by
\begin{align*}
&\vartheta_1 \left(\frac{n-\ell}{\pi r}; i \kappa \right) 
\vartheta_1 \left(\frac{m-k}{\pi r}; i \kappa \right)
\vartheta_1 \left( \frac{\alpha+m+k+n}{\pi r}; i \kappa \right)
\nonumber\\
& \quad \times
\vartheta_1 \left(\frac{\beta+2n+m}{\pi r}; i \kappa \right) 
\vartheta_1 \left(\frac{\alpha-\beta+m-n}{\pi r}; i \kappa \right),
\end{align*}
then we obtain the equation
\begin{align*}
&\vartheta_1 \left(\frac{n-\ell+m-k}{\pi r}; i \kappa \right) 
\vartheta_1 \left(\frac{\alpha+m+k+n}{\pi r}; i \kappa \right) 
\nonumber\\
&\quad \times
\vartheta_1 \left(\frac{\beta+n+\ell+m}{\pi r}; i \kappa \right) 
\vartheta_1 \left(\frac{\alpha-\beta+m-n}{\pi r}; i \kappa \right)
\nonumber\\
&\qquad
=\vartheta_1 \left(\frac{m-k}{\pi r}; i \kappa \right) 
\vartheta_1 \left(\frac{\alpha+m+k+\ell}{\pi r}; i \kappa \right)
\nonumber\\
&\qquad \quad \times
\vartheta_1 \left(\frac{\beta+2n+m}{\pi r}; i \kappa \right) 
\vartheta_1 \left(\frac{\alpha-\beta+m-\ell}{\pi r}; i \kappa \right)
\nonumber\\
&\qquad
+ \vartheta_1 \left(\frac{n-\ell}{\pi r}; i \kappa \right) 
\vartheta_1 \left(\frac{\alpha+n+2m}{\pi r}; i \kappa \right)
\nonumber\\
&\qquad \quad
\times
\vartheta_1 \left(\frac{\beta+n+\ell+k}{\pi r}; i \kappa \right) 
\vartheta_1 \left(\frac{\alpha-\beta+k-n}{\pi r}; i \kappa \right).
\end{align*}
This is a special case of the addition formula of theta functions (\ref{eqn:Riemann1})
with
\begin{align*}
&
x=\frac{1}{2 \pi r} (\alpha+2m+\ell), \quad
y=\frac{1}{2 \pi r} (\alpha+2m+2n-\ell),
\nonumber\\
&
u=\frac{1}{2 \pi r} (2\beta-\alpha+2n+\ell),
\quad
v=-\frac{1}{2 \pi r} (\alpha+\ell+2 k),
\end{align*}
and $\tau=i \kappa$,
since $\vartheta_1(-v; \tau)=-\vartheta_1(v; \tau)$.
Thus the proof is completed. \qed
\vskip 0.3cm

The weight (\ref{eqn:q_1}) for the elementary rightward step
$(t-1, x-1) \to (t,x)$ and
the transition weight (\ref{eqn:Q_2}) 
from $(s, x)$ to $(t, y)$ in the spatio-temporal plane (\ref{eqn:Lambda})
are obtained from the above as
\begin{align*}
q(t,x) &= w((t+x)/2, (t-x)/2),
\nonumber\\
Q(s, x; t, y) &=
w( \cP( ( (s+x)/2, (s-x)/2 ) \to ( (t+y)/2, (t-y)/2 ) ) ).
\end{align*}

\subsection{Excursion processes}
\label{sec:excursion}

Assume that
\begin{equation}
T \in \N.
\label{eqn:T1}
\end{equation}
Consider an excursion process on $\Z$
which starts from the origin 0 at time $t=0$
and returns to the origin 0 at time $t=2T$;
$X(t), t \in \{0,1, \dots, 2T\}$.
Denote the measure for trajectories
of this excursion process by $\P^{0,0}_{2T}$.
For an arbitrary integer $M \in \N$ and for an arbitrary
strictly increasing series of times $\{t_m \in \N_0 : m=0,1, \dots, M\}$ such that
\begin{equation}
0 =t_0 < t_1 < t_2 < \cdots < t_M \leq 2T, 
\label{eqn:times1}
\end{equation}
if
\begin{equation}
x^{(m)} \in \Z, \quad
t_m+x^{(m)} \in 2 \Z, 
\quad \mbox{for $m=1, 2, \dots, M$},
\label{eqn:conditionA1}
\end{equation}
then the multi-time joint measure
of $X(t), t \in \{0,1, \dots, 2T \}$ is given by
\begin{align}
& \P^{0,0}_{2T}(X(t_m)=x^{(m)}, m \in \{1,2, \dots, M\})
\nonumber\\
& \quad =
\prod_{m=1}^M
Q(t_{m-1}, x^{(m-1)}; t_m, x^{(m)})
\frac{Q(t_M, x^{(M)}; 2T, 0)}{Q(0, 0; 2T, 0)}.
\label{eqn:measure1}
\end{align}
If the condition (\ref{eqn:conditionA1}) is
not satisfied, then we assume
\begin{equation}
\P^{0,0}_{2T}(X(t_m)=x^{(m)}, m \in \{1,2, \dots, M\})=0.
\label{eqn:measure2}
\end{equation}
By the Chapman-Kolmogorov equation (\ref{eqn:CK1})
for $Q(s,x;t,y)$, 
\begin{align*}
&\sum_{x^{(m')} \in \Z} 
\P^{0,0}_{2T}(X(t_m)=x^{(m)}, m \in \{1,2, \dots, M\})
\nonumber\\
& \quad
=\P^{0,0}_{2T}(X(t_m)=x^{(m)}, m \in \{1,2, \dots, M\} \setminus \{m'\})
\end{align*}
for any $m' \in \{1,2, \dots, M\}$.
Therefore, we see that
\begin{align*}
&\sum_{x^{(2)} \in \Z}
\cdots \sum_{x^{(M)} \in \Z}
\P^{0,0}_{2T}(X(t_m)=x^{(m)}, m \in \{1,2, \dots, M\})
\nonumber\\
& \quad 
= \P^{0,0}_{2T}(X(t_1)=x^{(1)})
\nonumber\\
& \quad
= Q(0, 0; t_1, x^{(1)}) 
\frac{Q(t_1, x^{(1)}; 2T, 0)}
{Q(0, 0; 2T, 0)}.
\end{align*}
Again by the Chapman-Kolmogorov equation
(\ref{eqn:CK1}) for $Q(s,x;t,y)$, we have
\begin{equation}
\sum_{x^{(1)} \in \Z} \P^{0,0}_{2T}(X(t_1)=x^{(1)}) = 1.
\label{eqn:normalization1}
\end{equation}
That is, this measure $\P^{0,0}_{2T}$ is well normalized.
When $\alpha, \beta \in \R$, $r>0, \kappa >0$,
the measure $\P^{0,0}_{2T}(\cdot) = \P^{0,0}_{2T}(\cdot; r, \alpha, \beta, \kappa)$
is real-valued, but it is signed in general.

Excursion processes are also called
bridges, pinned processes, or tied-down processes.
See Part I. Section IV.4 in \cite{BS02}
for a Brownian bridge.

\subsection{Parameterization for excursion processes}
\label{sec:parameterization}

An explicit expression for the single-time measure
is obtained by applying (\ref{eqn:Q_2}) to the formula
\begin{equation}
\P^{0,0}_{2T}(X(t)=x)
=Q(0,0; t, x)
\frac{Q(t, x; 2T, 0)}{Q(0, 0; 2T, 0)},
\quad t \in \{0,1, \dots, 2T\}. 
\label{eqn:P1_0}
\end{equation}
We found that the following parameterization makes
the expression much simpler,
\begin{equation}
\alpha=\alpha_0-\frac{1}{2}(3T+1), \quad
\beta=\beta_0-\frac{1}{2}(3T+1),
\label{eqn:para2}
\end{equation}
where $\alpha_0$ and $\beta_0$ are still arbitrary.
Let
\begin{align}
\Lambda_{2T} &= \{ (t, x) \in \Lambda : 0 \leq t \leq T, -t \leq x \leq t \}
\nonumber\\
& \quad \cup \{(t, x) \in \Lambda : T+1 \leq t \leq 2T, t-2T \leq x \leq -t+2T \}.
\label{eqn:Lambda_2T_0}
\end{align}
The obtained expression is the following, 
\begin{align}
& \P^{0,0}_{2T}(X(t)=x) = \P^{0,0}_{2T}(X(t)=x; r, \alpha_0, \beta_0, \kappa) 
\nonumber\\
& \quad
= \begin{cases}
c_1(t; T) 
\vartheta_1 \left( \frac{\alpha_0-\beta_0-x}{\pi r}; i \kappa \right)
& \\
\displaystyle{
\quad \times 
\prod_{n=1}^{(t+x)/2} 
\frac{\vartheta_1 \left(\frac{2 \alpha_0-(2n+3T-2t-1)}{2 \pi r}; i \kappa \right)}
{\vartheta_1 \left( \frac{n}{\pi r}; i \kappa \right)
\vartheta_1 \left( \frac{\alpha_0-\beta_0-n}{\pi r}; i \kappa \right)}
\prod_{n=1}^{(t-x)/2} 
\frac{\vartheta_1 \left(\frac{2 \beta_0-(2n+3T-2t-1)}{2 \pi r}; i \kappa \right)}
{\vartheta_1 \left( \frac{n}{\pi r}; i \kappa \right)
\vartheta_1 \left( \frac{\alpha_0-\beta_0+n}{\pi r}; i \kappa \right)}
}
& \\
\displaystyle{
\quad \times 
\prod_{n=1}^{\{(2T-t)-x\}/2} 
\frac{\vartheta_1 \left(\frac{2 \alpha_0+(2n+2t-T-1)}{2 \pi r}; i \kappa \right)}
{\vartheta_1 \left( \frac{n}{\pi r}; i \kappa \right)
\vartheta_1 \left( \frac{\alpha_0-\beta_0+n}{\pi r}; i \kappa \right)}
\prod_{n=1}^{\{(2T-t)+x\}/2} 
\frac{\vartheta_1 \left(\frac{2 \beta_0+(2n+2t-T-1)}{2 \pi r}; i \kappa \right)}
{\vartheta_1 \left( \frac{n}{\pi r}; i \kappa \right)
\vartheta_1 \left( \frac{\alpha_0-\beta_0-n}{\pi r}; i \kappa \right)}
},
& \\
\hskip 8cm \mbox{if $(t, x) \in \Lambda_{2T}$},
& \\
0,
\hskip 7.6cm \mbox{otherwise}, 
&
\end{cases}
\label{eqn:P1_ell_1}
\end{align}
where
\begin{align}
c_1(t; T)
={}& 
\frac{
\prod_{n=1}^t \vartheta_1 \left( \frac{n}{\pi r} ; i \kappa \right) 
\prod_{n=1}^{2T-t} \vartheta_1 \left( \frac{n}{\pi r} ; i \kappa \right) 
}
{
\vartheta_1 \left( \frac{\alpha_0-\beta_0}{\pi r} ; i \kappa \right)
\prod_{n=1}^{2T} \vartheta_1 \left( \frac{n}{\pi r} ; i \kappa \right)
}
\nonumber\\
&\times
\prod_{n=1}^T
\frac{
\left\{ \vartheta_1 \left( \frac{n}{\pi r} ; i \kappa \right) \right\}^2
\vartheta_1 \left( \frac{\alpha_0-\beta_0-n}{\pi r} ; i \kappa \right)
\vartheta_1 \left( \frac{\alpha_0-\beta_0+n}{\pi r} ; i \kappa \right)
}
{
\vartheta_1\left( \frac{2 \alpha_0+(2n-T-1)}{2 \pi r} ; i \kappa \right)
\vartheta_1\left( \frac{2 \beta_0+(2n-T-1)}{2 \pi r} ; i \kappa \right)
}.
\label{eqn:P1_ell_c}
\end{align}

From now on, we assume the parameterization (\ref{eqn:para2})
with $\alpha_0, \beta_0 \in \R$, $r, \kappa >0$ for the
signed measure
$\P^{0,0}_{2T}(\cdot)=\P^{0,0}_{2T}(\cdot; r, \alpha_0, \beta_0, \kappa)$.

\subsection{Reduction to trigonometric and classical measures}
\label{sec:reduction}

By the form (\ref{eqn:q_1}) and 
the asymptotic property (\ref{eqn:limit_1}) of $\vartheta_1$,
\[
\widehat{q}(t,x;r,\alpha_0, \beta_0) \equiv 
\lim_{\kappa \to \infty} q(t, x; r, \alpha_0, \beta_0, \kappa)
\]
is well defined (see Section \ref{sec:positivity_tri}).
Let
$\widehat{\P}^{0,0}_{2T}(\cdot; r, \alpha_0, \beta_0)$
be the signed measure obtained from 
$\P^{0,0}_{2T}(\cdot; r, \alpha_0, \beta_0, \kappa)$
by taking the limit $\kappa \to \infty$.
For example, 
the single-time measure
(\ref{eqn:P1_ell_1}) gives
\begin{align}
& \widehat{\P}^{0,0}_{2T}(X(t)=x) = \widehat{\P}^{0,0}_{2T}(X(t)=x; r, \alpha_0, \beta_0) 
\nonumber\\
& \quad = \begin{cases}
\widehat{c}_1(t; T) 
\sin \left( \frac{\alpha_0-\beta_0-x}{r} \right)
& \\
\displaystyle{ \quad \times 
\prod_{n=1}^{(t+x)/2} 
\frac{\sin \left(\frac{2 \alpha_0-(2n+3T-2t-1)}{2r} \right)}
{\sin \left( \frac{n}{r} \right)
\sin \left( \frac{\alpha_0-\beta_0-n}{r} \right)}
\prod_{n=1}^{(t-x)/2} 
\frac{\sin \left(\frac{2 \beta_0-(2n+3T-2t-1)}{2r} \right)}
{\sin \left( \frac{n}{r} \right)
\sin \left( \frac{\alpha_0-\beta_0+n}{r} \right)}
} & \\
\displaystyle{
\quad \times 
\prod_{n=1}^{\{(2T-t)-x\}/2} 
\frac{\sin \left(\frac{2 \alpha_0+(2n+2t-T-1)}{2r} \right)}
{\sin \left( \frac{n}{r} \right)
\sin \left( \frac{\alpha_0-\beta_0+n}{r} \right)}
\prod_{n=1}^{\{(2T-t)+x\}/2} 
\frac{\sin \left(\frac{2 \beta_0+(2n+2t-T-1)}{2r} \right)}
{\sin \left( \frac{n}{r} \right)
\sin \left( \frac{\alpha_0-\beta_0-n}{r} \right)}
},
& \\
\hskip 8cm \mbox{if $(t, x) \in \Lambda_{2T}$}, 
& \\
0, 
\hskip 7.6cm \mbox{otherwise},
\end{cases}
\label{eqn:P1_tri1_1}
\end{align}
where
\begin{align}
\widehat{c}_1(t; T)
={}& 
\frac{
\prod_{n=1}^t \sin \left( \frac{n}{r}  \right) 
\prod_{n=1}^{2T-t} \sin \left( \frac{n}{r}  \right) 
}
{
\sin \left( \frac{\alpha_0-\beta_0}{r}  \right)
\prod_{n=1}^{2T} \sin \left( \frac{n}{r}  \right)
}
\nonumber\\
&\times
\prod_{n=1}^T
\frac{
\sin^2 \left( \frac{n}{r}  \right) 
\sin \left( \frac{\alpha_0-\beta_0-n}{r}  \right)
\sin \left( \frac{\alpha_0-\beta_0+n}{r}  \right)
}
{
\sin\left( \frac{2 \alpha_0+(2n-T-1)}{2r}  \right)
\sin\left( \frac{2 \beta_0+(2n-T-1)}{2r}  \right)
}.
\label{eqn:P1_tri1_c}
\end{align}

Now we consider the further limit.
We see that
\[
\widetilde{q}(t,x;r, \alpha_0)
\equiv \lim_{\widehat{\beta}_0 \to \infty}
\widehat{q}(t, x; r, \alpha_0, i \widehat{\beta}_0)
=\frac{\sin[\{2 \alpha_0 -(3T+1) +(3t-x)\}/2r]}
{\sin[\{2 \alpha_0-(3T+1)+(t+x)\}/2r]}.
\]
We write the corresponding signed measure as
$\widetilde{\P}^{0,0}_{2T}(\cdot)=\widetilde{\P}^{0,0}_{2T}(\cdot; r, \alpha_0)$.
In particular, (\ref{eqn:P1_tri1_1}) with (\ref{eqn:P1_tri1_c})
gives
\begin{align}
& \widetilde{\P}^{0,0}_{2T}(X(t)=x) = \widetilde{\P}^{0,0}_{2T}(X(t)=x; r, \alpha_0) 
\nonumber\\
& \quad = \begin{cases}
\displaystyle{
\widetilde{c}_1(t; T) 
\prod_{n=1}^{(t+x)/2} 
\frac{\sin \left(\frac{2 \alpha_0-(2n+3T-2t-1)}{2r} \right)}
{\sin \left( \frac{n}{r} \right)}
\prod_{n=1}^{(t-x)/2} 
\frac{1}
{\sin \left( \frac{n}{r} \right)}
}
& \\
\displaystyle{
\quad 
\times 
\prod_{n=1}^{\{(2T-t)-x\}/2} 
\frac{\sin \left(\frac{2 \alpha_0+(2n+2t-T-1)}{2r} \right)}
{\sin \left( \frac{n}{r} \right)}
\prod_{n=1}^{\{(2T-t)+x\}/2} 
\frac{1}
{\sin \left( \frac{n}{r} \right)}
}, 
& \quad \mbox{if $(t,x) \in \Lambda_{2T}$}, 
\\
0, & \quad \mbox{otherwise}, 
\end{cases}
\label{eqn:P1_tri2_1}
\end{align}
where
\begin{equation}
\widetilde{c}_1(t; T)
= 
\frac{
\prod_{n=1}^t \sin \left( \frac{n}{r} \right) 
\prod_{n=1}^{2T-t} \sin \left( \frac{n}{r} \right) 
}
{
\prod_{n=1}^{2T} \sin \left( \frac{n}{r} \right)
}
\prod_{n=1}^T
\frac{
\sin^2 \left( \frac{n}{r}  \right) 
}
{
\sin\left( \frac{2 \alpha_0+(2n-T-1)}{2r} \right)
}.
\label{eqn:P1_tri2_c}
\end{equation}

We can show that if we put $\alpha_0=i \widehat{\alpha}_0$
in (\ref{eqn:P1_tri1_1}) and (\ref{eqn:P1_tri1_c}) and
take the limit $\widehat{\alpha}_0 \to \infty$,
then we obtain the result
(\ref{eqn:P1_tri2_1}) and (\ref{eqn:P1_tri2_c}) with the
following replacement,
\[
\alpha_0 \to \beta_0, \quad
x \to -x.
\]

For the trigonometric measures
$\widehat{\P}^{0,0}_{2T}$ and $\widetilde{\P}^{0,0}_{2T}$,
if we take the further limit $r \to \infty$,
we will obtain a measure $\P^{0,0}_{2T,{\rm cl}}$
such that
\begin{equation}
\P^{0,0}_{2T, {\rm cl}}(X(t)=x)
= \begin{cases}
\frac{
\displaystyle{ \binom{t}{(t+x)/2} \binom{2T-t}{\{(2T-t)+x\}/2}
}}
{\displaystyle{ \binom{2T}{T}
}},
& \quad \mbox{if $(t, x) \in \Lambda_{2T}$},
\\
0, & \quad \mbox{otherwise}.
\end{cases}
\label{eqn:classical1}
\end{equation}
This gives the single-time probability measure
for the excursion process of
the {\it classical random walk} (that is, the simple and symmetric random walk)
on $\Z^2$ starting from 0 and returning to 0 with time duration $2T$.

\SSC
{Positivity Conditions for Measures} \label{sec:positivity}
\subsection{Elliptic excursion processes} 
\label{sec:positivity_elliptic}
\begin{figure}
\begin{center}
\includegraphics[width=0.5\textwidth]{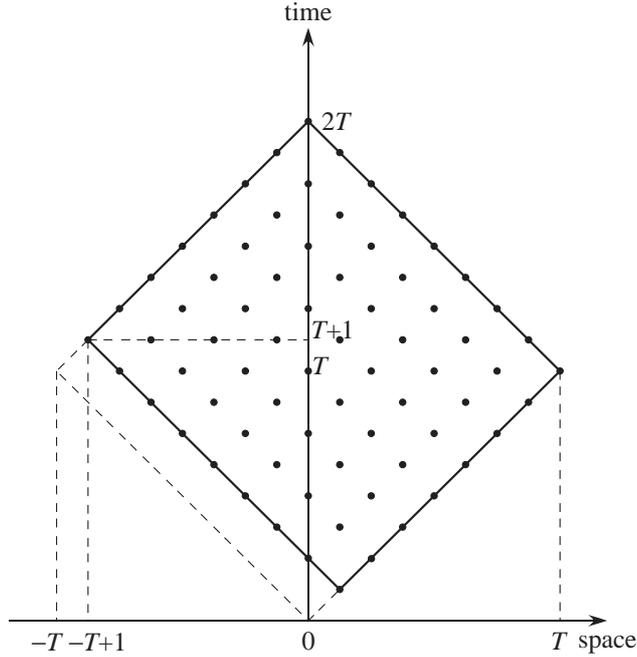}
\end{center}
\caption{The collection $\Lambda^{\ast}_{2T}$ 
of points $(t, x)$ such that
any trajectory of the excursion process from
0 to 0 with time duration $2T$ can contain
the rightward step $(t-1, x-1) \to (t, x)$.}
\label{fig:region}
\end{figure}
With the parameterization (\ref{eqn:para2}), 
the elementary weight-function $q(t, x)$ given by (\ref{eqn:q_1})
becomes the following,
\begin{equation}
q(t, x)=q(t, x; r, \alpha_0, \beta_0, \kappa)
= \prod_{j=1}^5 
\frac{\vartheta_1(\zeta_j/\pi; i \kappa)}
{\vartheta_1(\eta_j/\pi; i \kappa)}, 
\label{eqn:q_2}
\end{equation}
where
\begin{align}
\zeta_1 &= \frac{\alpha_0}{r} - \frac{3T+1}{2r} + \frac{3t-x}{2r},
\nonumber\\
\zeta_2 &= \frac{\beta_0}{r} - \frac{3T+1}{2r} + \frac{t+x}{r},
\qquad
\zeta_3 = \zeta_2 - \frac{1}{r},
\nonumber\\
\zeta_4 &= \frac{\alpha_0 - \beta_0}{r} - \frac{t+x}{2r},
\qquad
\zeta_5 = \zeta_4 + \frac{1}{r},
\label{eqn:zeta1}
\end{align}
and
\begin{align}
\eta_1 &= \frac{\alpha_0}{r} - \frac{3T+1}{2r} + \frac{t+x}{2r},
\nonumber\\
\eta_2 &= \frac{\beta_0}{r} - \frac{3T+1}{2r} + \frac{3t+x}{2r},
\qquad
\eta_3 = \eta_2 - \frac{1}{r},
\nonumber\\
\eta_4 &= \frac{\alpha_0-\beta_0}{r} - \frac{x}{r},
\qquad
\eta_5 = \eta_4+\frac{1}{r}.
\label{eqn:eta1}
\end{align}
On the spatio-temporal plane $\Lambda$, the collection 
of all points $(t, x)$ such that
any trajectory of the present excursion process from
0 to 0 with time duration $2T$ can contain
the rightward step $(t-1, x-1) \to (t, x)$ weighted with (\ref{eqn:q_2})
is given by
\begin{align}
\Lambda^{\ast}_{2T} &= \{ (t, x) \in \Lambda : 1 \leq t \leq T, -t+2 \leq x \leq t \}
\nonumber\\
& \quad \cup \{(t, x) \in \Lambda : T+1 \leq t \leq 2T, t-2T \leq x \leq -t+2T \}.
\label{eqn:Lambda_2T}
\end{align}
See Fig. \ref{fig:region}. 
If 
\begin{equation}
q(t, x) \geq 0, \quad \forall (t, x) \in \Lambda^{\ast}_{2T},
\label{eqn:Q_positivity}
\end{equation}
then the transition weight (\ref{eqn:Q_2}) is
non-negative, and hence the
multi-time joint measure (\ref{eqn:measure1}) 
is non-negative definite.
In this case $\P^{0,0}_{2T}$
gives a probability measure
and the excursion process
$(\{X(t)\}_{t \in \{0,1, \dots, 2T\}}, \P^{0,0}_{2T})$
is well-defined.

\begin{thm}
\label{thm:positivity_main}
Put
\begin{equation}
\lambda=\frac{3T-1}{2 \pi r}.
\label{eqn:lambda1}
\end{equation}
The following are sufficient conditions so that
$\P^{0,0}_{2T}$ is non-negative
and gives a probability measure
for the excursion process $X(t), t \in \{0,1,\dots, 2T\}$,
\begin{align}
& 0 \leq \lambda < \frac{1}{2},
\label{eqn:condB1}
\\
& \pi \lambda < \frac{\alpha_0}{r} < \pi (1-\lambda),
\label{eqn:condB2}
\\
& \pi \lambda < - \frac{\beta_0}{r} < \pi (1-\lambda),
\label{eqn:condB3}
\\
& \frac{\alpha_0-\beta_0}{r}
< \pi \left\{ 1- \frac{2(T+1)}{3T-1} \lambda \right\}.
\label{eqn:condB4}
\end{align}
\end{thm}
\noindent{\it Proof.} \,
In the region $\Lambda^{\ast}_{2T}, T \in \N$, given by (\ref{eqn:Lambda_2T}),
the following inequalities hold,
\begin{equation}
1 \leq t \leq T, \quad x \geq -t+2, \quad x \leq t,
\label{eqn:ineqA1}
\end{equation}
or
\begin{equation}
T+1 \leq t \leq 2T, \quad x \geq t-2T, \quad x \leq -t+2T.
\label{eqn:ineqA2}
\end{equation}
When the inequalities (\ref{eqn:ineqA1}) hold,
\begin{align}
-t \leq -x \leq t-2 
\quad &\Rightarrow \quad 
2t \leq 3t-x \leq 4t-2 
\nonumber\\
&\Rightarrow \quad
2 \leq 3t-x \leq 4T-2,
\nonumber
\\
2 \leq t+x \leq 2t 
\quad &\Rightarrow \quad 2 \leq t+x \leq 2T.
\label{eqn:ineqA4}
\end{align}
By the first inequality of (\ref{eqn:ineqA1}), $2 \leq 2t \leq 2T$.
Combining this with (\ref{eqn:ineqA4}), we have
\begin{equation}
4 \leq 3t+x \leq 4T.
\label{eqn:ineqA5}
\end{equation}
Similarly, when the inequalities (\ref{eqn:ineqA2}) hold, 
we have
\begin{equation}
6 \leq 3t-x \leq 6T, \quad
2 \leq t+x \leq 2T, \quad
2T+4 \leq 3t+x \leq 6T.
\label{eqn:ineqA6}
\end{equation}
Therefore, the following inequalities hold in $\Lambda^{\ast}_{2T}, T \in \N$,
\begin{equation}
2 \leq 3t-x \leq 6T, \quad
2 \leq t+x \leq 2T, \quad
4 \leq 3t+x \leq 6T.
\label{eqn:ineqB1}
\end{equation}
By the first and the second inequalities in (\ref{eqn:ineqB1}),
we have
\begin{align*}
\frac{\alpha_0}{r}-\frac{3T-1}{2r} 
&\leq \zeta_1 
\leq \frac{\alpha_0}{r}+\frac{3T-1}{2r},
\nonumber\\
\frac{\alpha_0}{r}-\frac{3T-1}{2r} 
&\leq \eta_1
\leq \frac{\alpha_0}{r}+\frac{3T-1}{2r}-\frac{2T}{r}.
\end{align*}
Hence, if
\begin{align}
& 0 < \frac{\alpha_0}{r}-\frac{3T-1}{2r} \quad
\mbox{and} \quad
\frac{\alpha_0}{r}+ \frac{3T-1}{2r} < \pi
\nonumber\\
&\Longleftrightarrow \quad
\frac{3T-1}{2r} < \frac{\alpha_0}{r} 
< \pi - \frac{3T-1}{2r},
\label{eqn:ineqB2}
\end{align}
then $0< \zeta_1 < \pi$, $0 < \eta_1 < \pi$,
and 
\begin{equation}
0 < \frac{\vartheta_1(\zeta_1/\pi; i \kappa)}
{\vartheta_1(\eta_1/\pi; i \kappa)} < \infty.
\label{eqn:ineqB3}
\end{equation}
Similarly, by the second and the third inequalities in (\ref{eqn:ineqB1}),
we have
\begin{align*}
\frac{\beta_0}{r}-\frac{3T-1}{2r} +\frac{1}{r}
&\leq \zeta_2 
\leq \frac{\beta_0}{r}+\frac{3T-1}{2r}-\frac{T}{r},
\nonumber\\
\frac{\beta_0}{r}-\frac{3T-1}{2r} 
&\leq \zeta_3
\leq \frac{\beta_0}{r}+\frac{3T-1}{2r}-\frac{T-1}{r},
\nonumber\\
\frac{\beta_0}{r}-\frac{3T-1}{2r} +\frac{1}{r}
&\leq \eta_2
\leq \frac{\beta_0}{r}+\frac{3T-1}{2r},
\nonumber\\
\frac{\beta_0}{r}-\frac{3T-1}{2r}
&\leq \eta_3
\leq \frac{\beta_0}{r}+\frac{3T-1}{2r}-\frac{1}{r}.
\end{align*}
Hence, if
\begin{align}
& -\pi < \frac{\beta_0}{r}-\frac{3T-1}{2r} \quad
\mbox{and} \quad
\frac{\beta_0}{r}+ \frac{3T-1}{2r} < 0
\nonumber\\
&\Longleftrightarrow \quad
\frac{3T-1}{2r} < -\frac{\beta_0}{r} 
< \pi - \frac{3T-1}{2r},
\label{eqn:ineqB4}
\end{align}
then $-\pi < \zeta_j < 0$, $-\pi < \eta_j < 0$ for $j=2,3$
and 
\begin{equation}
0 < \frac{\vartheta_1(\zeta_2/\pi; i \kappa) \vartheta_1(\zeta_3/\pi; i \kappa)}
{\vartheta_1(\eta_2/\pi; i \kappa) \vartheta_1(\eta_3/\pi; i \kappa)} < \infty.
\label{eqn:ineqB5}
\end{equation}
By the second inequality in (\ref{eqn:ineqB1})
and the inequality $-T \leq x \leq T$ satisfied in $\Lambda^{\ast}_{2T}, T \in \N$, 
we have
\begin{align*}
\frac{\alpha_0-\beta_0}{r}- \frac{T}{r}
&\leq \zeta_4 \leq \frac{\alpha_0-\beta_0}{r}-\frac{1}{r},
\nonumber\\
\frac{\alpha_0-\beta_0}{r}- \frac{T}{r} +\frac{1}{r}
&\leq \zeta_5 \leq \frac{\alpha_0-\beta_0}{r},
\nonumber\\
\frac{\alpha_0-\beta_0}{r}- \frac{T}{r}
&\leq \eta_4 \leq \frac{\alpha_0-\beta_0}{r}+\frac{T}{r},
\nonumber\\
\frac{\alpha_0-\beta_0}{r}- \frac{T}{r} +\frac{1}{r}
&\leq \eta_5 \leq \frac{\alpha_0-\beta_0}{r}+\frac{T+1}{r}.
\end{align*}
Hence, if
\begin{align}
& 0 < \frac{\alpha_0-\beta_0}{r}-\frac{T}{r} \quad
\mbox{and} \quad
\frac{\alpha_0-\beta_0}{r}+ \frac{T+1}{r} < \pi
\nonumber\\
&\Longleftrightarrow \quad
\frac{T}{r} < \frac{\alpha_0-\beta_0}{r} 
< \pi - \frac{T+1}{r},
\label{eqn:ineqB6}
\end{align}
then $0 < \zeta_j < \pi$, $0< \eta_j < \pi$ for $j=4,5$
and 
\begin{equation}
0 < \frac{\vartheta_1(\zeta_4/\pi; i \kappa) \vartheta_1(\zeta_5/\pi; i \kappa)}
{\vartheta_1(\eta_4/\pi; i \kappa) \vartheta_1(\eta_5/\pi; i \kappa)} < \infty.
\label{eqn:ineqB7}
\end{equation}
Using $\lambda$ defined by (\ref{eqn:lambda1}),
the conditions (\ref{eqn:ineqB2}) and (\ref{eqn:ineqB4})
are expressed as (\ref{eqn:condB1})-(\ref{eqn:condB3}).
If we combine (\ref{eqn:ineqB2}) and (\ref{eqn:ineqB4}),
we obtain
\[
\frac{3T-1}{r} < \frac{\alpha_0-\beta_0}{r}
< 2 \pi - \frac{3T-1}{r}.
\]
For $T \in \N$, $T/r < (3T-1)/r$.
We can also verify that
if $\lambda<1/2$, then
\[
2\pi - \frac{3T-1}{r} 
> \pi - \frac{T+1}{r}
=\pi \left\{ 1- \frac{2(T+1)}{3T-1} \lambda \right\}.
\]
Therefore, if (\ref{eqn:condB4}) is satisfied
in addition to (\ref{eqn:condB1})-(\ref{eqn:condB3}),
(\ref{eqn:ineqB6}) holds.
Thus (\ref{eqn:ineqB3}), (\ref{eqn:ineqB5}),
and (\ref{eqn:ineqB7}) imply that
the elementary weight-function $q(t,x)$ given by
(\ref{eqn:q_2}) satisfies (\ref{eqn:Q_positivity}). 
The proof is complete. \qed
\vskip 0.3cm

We call $(\{X(t)\}_{t \in \{0,1,\dots, 2T\}}, \P^{0,0}_{2T})$
with the parameters satisfying the conditions
(\ref{eqn:condB1})-(\ref{eqn:condB4})
the {\it elliptic excursion process}
with time duration $2T$.

As implied by the above proof, $\alpha_0$ and $\beta_0$
can be interchanged in (\ref{eqn:condB2})-(\ref{eqn:condB4})
to obtain another set of sufficient conditions
for non-negative $\P^{0,0}_{2T}$, 
which results in reflection of excursion trajectories
as $x \leftrightarrow -x$.

\subsection{Trigonometric excursion processes} 
\label{sec:positivity_tri}

By (\ref{eqn:limit_1}), we see that
\begin{align}
\widehat{q}(t, x) 
&=\widehat{q}(t,x; r, \alpha_0, \beta_0)
\nonumber\\
&\equiv \lim_{\kappa \to \infty} q(t, x; r, \alpha_0, \beta_0, i \kappa)
= \prod_{j=1}^5 \frac{\sin \zeta_j}{\sin \eta_j},
\label{eqn:q_hat1}
\end{align}
with (\ref{eqn:zeta1}) and (\ref{eqn:eta1}). 
If $0 \leq \zeta \leq \pi$, then
$0 \leq \vartheta_1(\zeta/\pi; i \kappa) < \infty$
for $0 < \kappa < \infty$ and $0 \leq \sin \zeta \leq 1$,
while if $-\pi \leq \zeta \leq 0$,
then $0 \leq \vartheta_1(\zeta/\pi; i \kappa) < \infty$
for $0 < \kappa < \infty$ and $0 \leq \sin \zeta \leq 1$.
Therefore, the proof of Theorem \ref{thm:positivity_main} implies 
the following corollary.

\begin{cor}
\label{thm:positivity_tri}
If the conditions {\rm (\ref{eqn:condB1})-(\ref{eqn:condB4})} are satisfied, then
$\widehat{\P}^{0,0}_{2T}$ gives a probability measure
for the excursion process $X(t), t \in \{0,1,\dots, 2T\}$.
\end{cor}

We call $(\{X(t)\}_{t \in \{0,1,\dots, 2T\}}, \widehat{\P}^{0,0}_{2T})$
with the parameters satisfying the conditions
(\ref{eqn:condB1})-(\ref{eqn:condB4})
the {\it trigonometric excursion process}
with time duration $2T$.

\subsection{Simplified trigonometric excursion processes} 
\label{sec:positivity_simp_tri}

For (\ref{eqn:q_hat1}), we see that
\begin{align}
\widetilde{q}(t, x) &=
\widetilde{q}(t,x; r, \alpha_0)
\nonumber\\
&\equiv \lim_{\widehat{\beta}_0 \to \infty}
\widehat{q}(t, x; t, \alpha_0, i \widehat{\beta}_0)
= \frac{\sin \zeta_1}{\sin \eta_1},
\label{eqn:q_tilde1}
\end{align}
where $\zeta_1$ and $\eta_1$ are given by
(\ref{eqn:zeta1}) and (\ref{eqn:eta1}). 

\begin{cor}
\label{thm:positivity_simp_tri}
Assume that the conditions {\rm (\ref{eqn:condB1})} and {\rm (\ref{eqn:condB2})}
are satisfied. 
Then $\widetilde{\P}^{0,0}_{2T}$ gives a probability measure
for the excursion process $X(t), t \in \{0,1,\dots, 2T\}$.
\end{cor}

We call $(\{X(t)\}_{t \in \{0,1,\dots, 2T\}}, \widetilde{\P}^{0,0}_{2T})$
with the parameters satisfying the conditions
(\ref{eqn:condB1}) and (\ref{eqn:condB2})
the {\it simplified trigonometric excursion process} 
with time duration $2T$.

\SSC
{Numerical Study of Trajectories} \label{sec:trajectories}

As shown at the end of Section \ref{sec:reduction},
if we take the limit $r \to \infty$ with a fixed finite $T<\infty$,
the present process is reduced to the excursion process
of a classical random walk,
whose trajectory is distributed according to (\ref{eqn:classical1}).
Here we consider the case such that
the long-term limit $T \to \infty$ is taken as well as $r \to \infty$.
Using Stirling's formula,
$n! \sim \sqrt{2 \pi n} n^n e^{-n}$ in $n \to \infty$,
it is easy to verify that
\begin{align}
& \lim_{T \to \infty} \frac{\sqrt{T}}{2}
\P^{0,0}_{2 T, {\rm cl}}
(X(Ts)=\sqrt{T} \xi)
\nonumber\\
& \quad
= \frac{1}{\sqrt{\pi s(2-s)}}
e^{-\xi^2/\{s(2-s)\}},
\quad s \in (0, 2), \quad \xi \in \R.
\label{eqn:BM1}
\end{align}
This is nothing but the probability density
at time $s$ of the Brownian bridge
(see, for instance, Part I. Section IV.4 in \cite{BS02})
starting from 0 and returning to 0 with time duration $2$.

In this section, we report the numerical study
of the long-term behavior in $T \to \infty$ of the
present elliptic and trigonometric excursion processes.
In order to realize any non-classical behavior in $T \to \infty$,
the parameter $\lambda$ introduced as (\ref{eqn:lambda1})
should be non-zero.
It implies that we have to take the double limit
$r \to \infty, T \to \infty$ in which
$r$ should be proportional to $T$ 
at least asymptotically in this limit.
In this section, we assume a proportional relation
\begin{equation}
\pi r = \sigma T,
\label{eqn:sigma1}
\end{equation}
with a fixed factor $\sigma$.
Since
\begin{equation}
\lambda < \frac{3T}{2 \pi r} = \frac{3}{2 \sigma},
\label{eqn:sigma2}
\end{equation}
the condition (\ref{eqn:condB1}) 
of Theorem \ref{thm:positivity_main} is satisfied, if
\begin{equation}
\sigma \geq 3.
\label{eqn:condC1}
\end{equation}
Moreover, by (\ref{eqn:sigma2}) and
the fact that $1/3 < (T+1)/(3T-1) \leq 1$
for $T \in \N$, we can verify that
if
\begin{align}
\frac{3}{2 \sigma} \pi &\leq \frac{\alpha_0}{r} \leq 
\left( 1 -\frac{3}{2 \sigma} \right) \pi,
\label{eqn:condC2}
\\
\frac{3}{2 \sigma} \pi &\leq - \frac{\beta_0}{r} \leq 
\left( 1 -\frac{3}{2 \sigma} \right) \pi,
\label{eqn:condC3}
\\
\frac{\alpha_0-\beta_0}{r}
&\leq \left( 1-\frac{3}{\sigma} \right) \pi,
\label{eqn:condC4}
\end{align}
then the conditions (\ref{eqn:condB2})-(\ref{eqn:condB4})
of Theorem \ref{thm:positivity_main} are satisfied.

\subsection{Simplified trigonometric excursion processes} 
\label{sec:numerical_simp_tri}
\begin{figure}
\begin{center}
\includegraphics[width=0.5\textwidth]{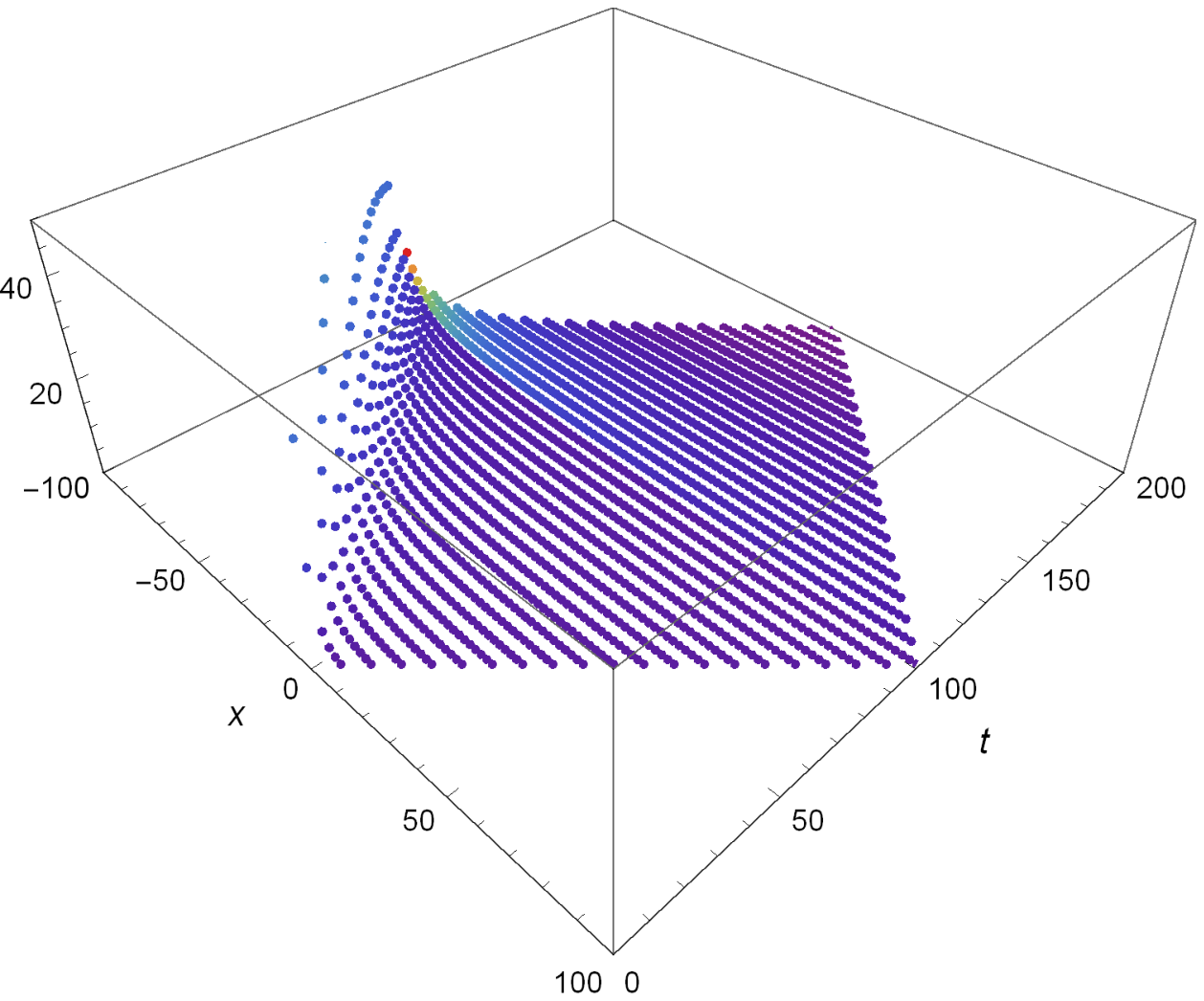}
\end{center}
\caption{
The elementary weight $\widetilde{q}(t,x)$ is shown as a function
of $(t, x)$ in $\Lambda_{2T}^{\ast}$ for $T=100$. 
The dependence on $(t,x)$ is very simple. 
}
\label{fig:simp_tri_q}
\end{figure}

The conditions of Corollary \ref{thm:positivity_simp_tri} are
satisfied, if (\ref{eqn:condC1}) and (\ref{eqn:condC2}) are valid.
As mentioned above, the parameter $\lambda$ will represent
the degree of deviation from the classical processes;
the $\lambda \to 0$ limit corresponds to the classical cases.
Here we will concentrate on the case in which
$\lambda$ is maximized, that is, the parameter $\sigma$
takes its minimal value $\sigma=3$.
Then the condition (\ref{eqn:condC2}) determines the value of $\alpha_0/r$
uniquely as follows,
\begin{equation}
\sigma=3, \quad \frac{\alpha_0}{r}=\frac{\pi}{2}.
\label{eqn:special1}
\end{equation}
Figure \ref{fig:simp_tri_q} shows 
the elementary weight $\widetilde{q}(t,x)$ as a function
of $(t, x)$ in $\Lambda_{2T}^{\ast}$ for $T=100$. 
The dependence on $(t,x)$ is very simple. 
In this special parameterization, (\ref{eqn:P1_tri2_1})
and (\ref{eqn:P1_tri2_c}) are written as follows,
\begin{align}
& \widetilde{\P}^{0,0}_{2T}(X(t)=x) = \widetilde{\P}^{0,0}_{2T}(X(t)=x; r, \alpha_0) 
\Big|_{\pi r= 3T, \alpha_0/r=\pi/2}
\nonumber\\
&\quad = \begin{cases} 
\displaystyle{
\widetilde{c}_2(t; T) 
\prod_{n=1}^{(t+x)/2} 
\frac{\cos \left(\frac{(2n+3T-2t-1) \pi}{6T} \right)}
{\sin \left( \frac{n \pi}{3T} \right)}
\prod_{n=1}^{(t-x)/2} 
\frac{1}
{\sin \left( \frac{n \pi}{3T} \right)}
} & \\
\displaystyle{
\quad \times 
\prod_{n=1}^{\{(2T-t)-x\}/2} 
\frac{\cos \left(\frac{(2n+2t-T-1) \pi}{6T} \right)}
{\sin \left( \frac{n \pi}{3T} \right)}
\prod_{n=1}^{\{(2T-t)+x\}/2} 
\frac{1}
{\sin \left( \frac{n \pi}{3T} \right)}
},
\quad & \mbox{if $(t, x) \in \Lambda_{2T}$},
\\
0, \quad & \mbox{otherwise},
\end{cases}
\label{eqn:P1_tri2_2}
\end{align}
where
\begin{equation}
\widetilde{c}_2(t; T)
= 
\frac{
\prod_{n=1}^t \sin \left( \frac{n \pi}{3T} \right) 
\prod_{n=1}^{2T-t} \sin \left( \frac{n \pi }{3T} \right) 
}
{
\prod_{n=1}^{2T} \sin \left( \frac{n \pi}{3T} \right)
}
\prod_{n=1}^T
\frac{
\sin^2 \left( \frac{n \pi}{3T}  \right) 
}
{
\cos \left( \frac{(2n-T-1) \pi}{6T} \right)
}.
\label{eqn:P1_tri2_c2}
\end{equation}
By definition of excursion processes
\[
\widetilde{\P}^{0,0}_{2T}(X(0)=x)
=\widetilde{\P}^{0,0}_{2T}(X(2T)=x)=\delta_{x 0}.
\]
In the expression (\ref{eqn:P1_tri2_2}) it is obvious to see
the following symmetry,
\begin{equation}
\widetilde{\P}^{0,0}_{2T}(X(t)=x)= \widetilde{\P}^{0,0}_{2T}(X(2T-t)=-x),
\quad 0 \leq t \leq 2T, \quad x \in \Z.
\label{eqn:symmetry1}
\end{equation}
Hence, at $t=T$ the distribution should be symmetric,
\[
\widetilde{\P}^{0,0}_{2T}(X(T)=x)= \widetilde{\P}^{0,0}_{2T}(X(T)=-x), \quad x \in \Z.
\]

\begin{figure}
\begin{center}
\includegraphics[width=0.5\textwidth]{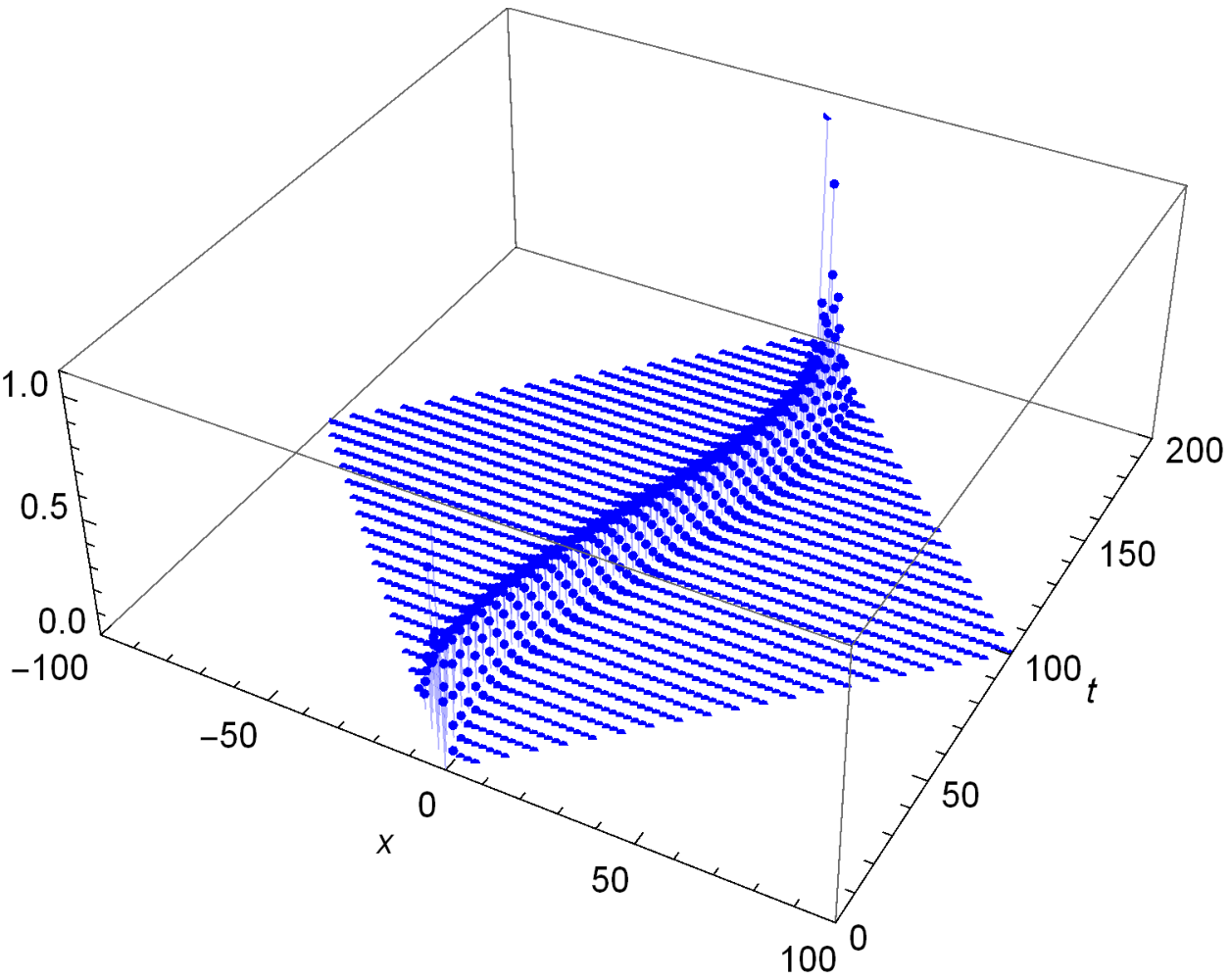}
\end{center}
\caption{Time-evolution of the probability distribution
$\widetilde{\P}^{0,0}_{2T}(X(t)=x)$ 
of the simplified trigonometric excursion process with $T=100$.
The maximum likelihood trajectory is reverse $S$-shaped. 
}
\label{fig:simp_tri_T=100}
\end{figure}

Figure \ref{fig:simp_tri_T=100} shows the time-dependence
of $\widetilde{\P}^{0,0}_{2T}(X(t)=x)$ 
given by (\ref{eqn:P1_tri2_2})
and (\ref{eqn:P1_tri2_c2})
for $T=100$. 
This figure clarifies that at each time $t \in \{0,1,\dots, 2T\}$
there is a single peak at a point 
which we denote by 
$\widetilde{x}^{\rm max}_{2T}(t)$,
and
\begin{align*}
& \widetilde{x}^{\rm max}_{2T}(t) < 0 \quad \mbox{for $1 \leq t \leq T-1$},
\nonumber\\
& \widetilde{x}^{\rm max}_{2T}(t) =0 \quad \mbox{at $t=T$},
\nonumber\\
& \widetilde{x}^{\rm max}_{2T}(t) > 0 \quad \mbox{for $T+1 \leq t \leq 2T-1$}.
\label{eqn:simp_tri_max_T=100}
\end{align*}
The line
$\{ \widetilde{x}^{\rm max}_{2T}(t) : 0 \leq t \leq 2T \}$ 
expresses the maximum likelihood trajectory
of the simplified trigonometric
excursion process.
As shown by Fig. \ref{fig:simp_tri_T=100},
it is reverse $S$-shaped.

\begin{figure}
\begin{center}
\includegraphics[width=0.5\textwidth]{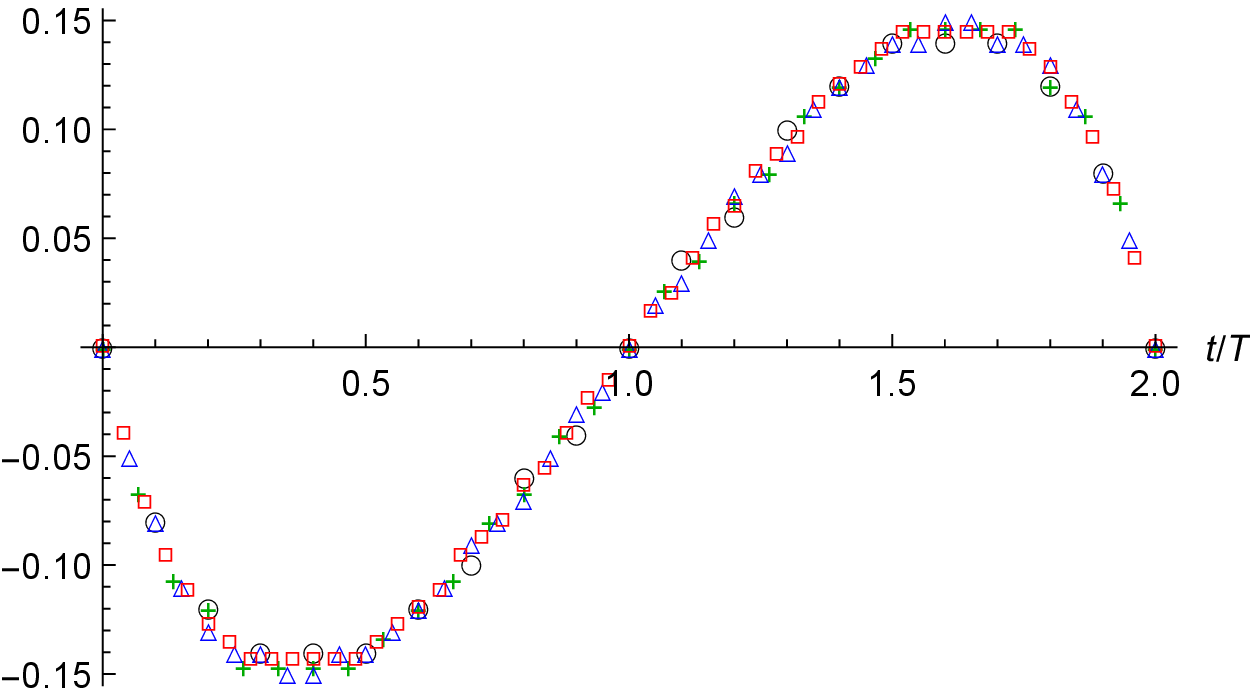}
\end{center}
\caption{
Numerical values of $\widetilde{x}^{\rm max}_{2T}(t)/T$, 
$t \in \{0,1, \dots, 2T\}$ are plotted for various time durations, 
$T=100$ (circles $\bigcirc$), 150 (crosses $+$), 
200 (triangles $\bigtriangleup$), 250 (squares $\Box$).
The observed data-collapse suggests the existence of a limit
curve $\widetilde{v}(t), t \in [0, 2]$ 
for the simplified trigonometric excursion process 
in the long-term limit $T \to \infty$.
}
\label{fig:simp_tri_max}
\end{figure}

Figure \ref{fig:simp_tri_max} shows
$\{ 
\widetilde{x}^{\rm max}_{2T}(sT)/T : 0 \leq s \leq 2 \}$
for $T=100, 150, 200, 250$.
We find accumulation of points with 
different values of $T$ into a curve,
which is called {\it data-collapse} in the scaling plots
studied in statistical mechanics.
It suggests that the values of
$\widetilde{x}^{\rm max}_{2T}(t)$ are proportional to $T$
in $T \to \infty$ and thus the
following scaled limit trajectory exists,
\[
\{ \widetilde{v}(s) : s \in [0, 2] \}
= \lim_{T \to \infty}
\left\{ 
\frac{ \widetilde{x}^{\rm max}_{2T}(sT)}{T} : 0 \leq s \leq 2
\right\},
\]
and it exhibits a nontrivial curve.
Due to (\ref{eqn:symmetry1}), it has the following
symmetry,
\[
\widetilde{v}(t)=- \widetilde{v} (2-t),
\quad t \in [0, 2],
\]
which proves $\widetilde{v}(1)=0$.

\subsection{Trigonometric excursion processes} 
\label{sec:numerical_tri}

\begin{figure}
\begin{center}
\includegraphics[width=0.5\textwidth]{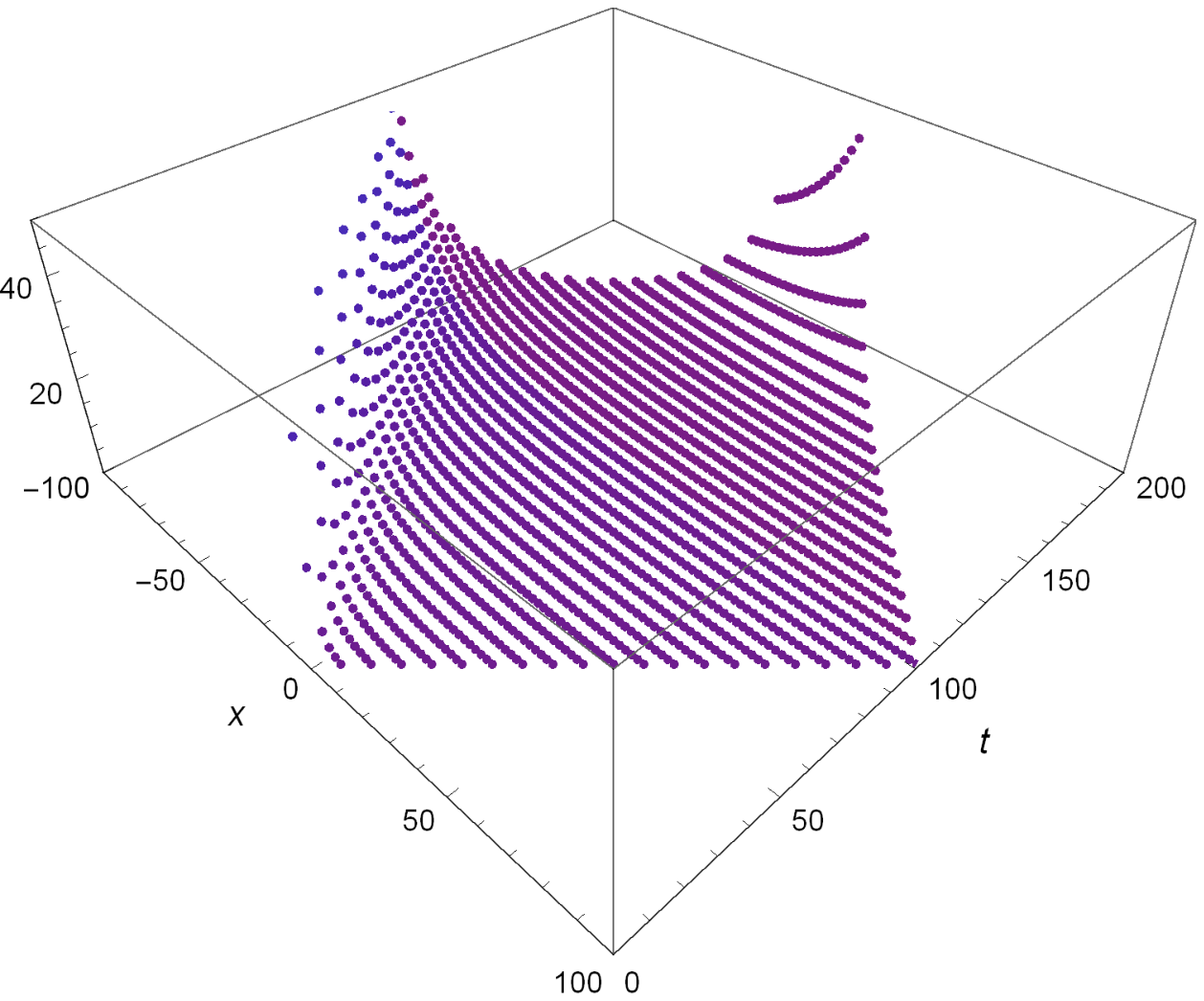}
\end{center}
\caption{
The elementary weight $\widehat{q}(t,x)$ is shown as a function
of $(t, x)$ in $\Lambda_{2T}^{\ast}$ for $T=100$. 
The dependence on $(t,x)$ is very simple. 
}
\label{fig:tri_q}
\end{figure}

For the trigonometric excursion process
$(\{X(t)\}_{t \in \{0,1,\dots, 2T\}}, \widehat{\P}^{0,0}_{2T})$,
we have to choose the parameters satisfying the
conditions (\ref{eqn:condC2})-(\ref{eqn:condC4}).
Since the combination of (\ref{eqn:condC2}) and (\ref{eqn:condC3})
gives $3 \pi/\sigma \leq (\alpha_0-\beta_0)/r$,
in order to satisfy also (\ref{eqn:condC4}), we
have to consider the case
\[
\frac{3}{\sigma} \pi \leq \left(1 -\frac{3}{\sigma} \right) \pi
\quad \Longleftrightarrow \quad
\sigma \geq 6.
\]
Here we choose the smallest value $\sigma = 6$
to make $\lambda$ be the largest.
In this case (\ref{eqn:condC2})-(\ref{eqn:condC4}) determine
the other parameters uniquely as
\begin{equation}
\sigma=6, \quad \frac{\alpha_0}{r}= -\frac{\beta_0}{r}=\frac{\pi}{4}.
\label{eqn:special2}
\end{equation}
Figure \ref{fig:tri_q} shows 
the elementary weight $\widehat{q}(t,x)$ as a function
of $(t, x)$ in $\Lambda_{2T}^{\ast}$ for $T=100$. 
The dependence on $(t,x)$ is very simple. 
In this special parameterization, (\ref{eqn:P1_tri1_1})
and (\ref{eqn:P1_tri1_c}) are written as follows,
\begin{align}
& \widehat{\P}^{0,0}_{2T}(X(t)=x) = 
\widehat{\P}^{0,0}_{2T}(X(t)=x; r, \alpha_0, \beta_0) \Big|_{\pi r= 6T, \alpha_0/r=\pi/4, \beta_0/r=-\pi/4}
\nonumber\\
& \quad 
= \begin{cases}
\widehat{c}_2(t; T) 
\cos \left( \frac{x \pi}{6T} \right)
& \\
\displaystyle{
\quad \times 
\prod_{n=1}^{(t+x)/2} 
\frac{\sin \left(  \frac{\pi}{4} -\frac{(2n+3T-2t-1) \pi}{12T} \right)}
{\sin \left( \frac{n \pi}{3T} \right)}
\prod_{n=1}^{(t-x)/2} 
\frac{\sin \left(\frac{\pi}{4} + \frac{(2n+3T-2t-1) \pi}{12T} \right)}
{\sin \left( \frac{n \pi}{3T} \right)}
} & \\
 \displaystyle{
\quad \times 
\prod_{n=1}^{\{(2T-t)-x\}/2} 
\frac{\sin \left( \frac{\pi}{4} + \frac{(2n+2t-T-1) \pi}{12T} \right)}
{\sin \left( \frac{n \pi}{3T} \right)}
\prod_{n=1}^{\{(2T-t)+x\}/2} 
\frac{\sin \left(\frac{\pi}{4} - \frac{(2n+2t-T-1) \pi}{12T} \right)}
{\sin \left( \frac{n \pi}{3T} \right)}
}, & \\
\hskip 8cm \mbox{if $(t, x) \in \Lambda_{2T}$},
& \\
0, \hskip 7.6cm \mbox{otherwise},
\end{cases}
\label{eqn:P1_tri1_2}
\end{align}
where
\begin{align}
\widehat{c}_2(t; T)
={}& 
\frac{ 2^T
\prod_{n=1}^t \sin \left( \frac{n \pi}{6T}  \right) 
\prod_{n=1}^{2T-t} \sin \left( \frac{n \pi}{6T}  \right) 
}
{
\prod_{n=1}^{2T} \sin \left( \frac{n \pi}{6T}  \right)
}
\prod_{n=1}^T
\frac{
\sin^2 \left( \frac{n \pi}{3T}  \right)
}
{
\cos \left( \frac{(2n-T-1) \pi}{6T}  \right)
}.
\label{eqn:P1_tri1_c2}
\end{align}
By definition of excursion process
\[
\widehat{\P}^{0,0}_{2T}(X(0)=x)
=\widehat{\P}^{0,0}_{2T}(X(2T)=x)=\delta_{x 0}.
\]
In the above expression, it is obvious to see
the following symmetry,
\begin{equation}
\widehat{\P}^{0,0}_{2T}(X(t)=x)= \widehat{\P}^{0,0}_{2T}(X(2T-t)=x),
\quad 0 \leq t \leq 2T, \quad x \in \Z.
\label{eqn:symmetry2}
\end{equation}

\begin{figure}
\begin{center}
\includegraphics[width=0.5\textwidth]{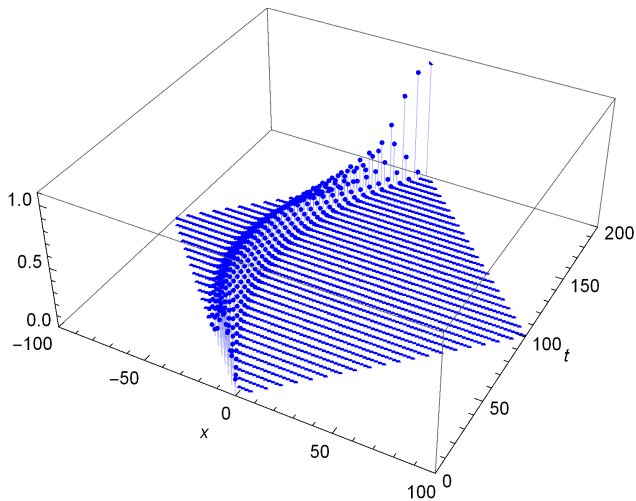}
\end{center}
\caption{Time-evolution of the probability distribution
$\widehat{\P}^{0,0}_{2T}(X(t)=x)$ 
of the trigonometric excursion process with $T=100$.
The maximum likelihood trajectory is $C$-shaped. }
\label{fig:tri_T=100}
\end{figure}
Figure \ref{fig:tri_T=100} shows the time-dependence
of $\widehat{\P}^{0,0}_{2T}(X(t)=x)$ 
given by (\ref{eqn:P1_tri1_2})
and (\ref{eqn:P1_tri1_c2})
for $T=100$. 
This figure suggests that at each time $t \in \{0,1,\dots, 2T\}$
there is a single peak at $x=\widehat{x}^{\rm max}_{2T}(t)$,
and
\[
\widehat{x}^{\rm max}_{2T}(t) < 0 \quad \mbox{for $1 \leq t \leq 2T-1$}.
\]
The line
$\{ \widehat{x}^{\rm max}_{2T}(t) : 0 \leq t \leq 2T\}$ 
expresses the maximum likelihood trajectory
of the trigonometric excursion process.
As shown by Fig. \ref{fig:tri_T=100}, it is $C$-shaped.

\begin{figure}
\begin{center}
\includegraphics[width=0.5\textwidth]{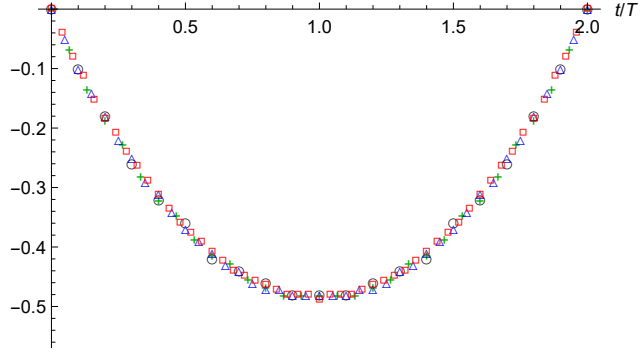}
\end{center}
\caption{
Numerical values of $\widehat{x}^{\rm max}_{2T}(t)/T$, 
$t \in \{0,1, \dots, 2T\}$ are plotted for various time durations, 
$T=100$ (circles $\bigcirc$), 150 (crosses $+$), 
200 (triangles $\bigtriangleup$), 250 (squares $\Box$).
The observed data-collapse suggests the existence of a limit
curve $\widehat{v}(t), t \in [0, 2]$ 
for the trigonometric excursion process 
in the long-term limit $T \to \infty$.
}
\label{fig:tri_max}
\end{figure}

Figure \ref{fig:tri_max} shows
$\{
\widehat{x}^{\rm max}_{2T}(sT)/T : 0 \leq s \leq 2
\}$
for $T=100, 150, 200, 250$.
The data-collapse is observed and it suggests that the values of
$\widehat{x}^{\rm max}_{2T}(t)$ are proportional to $T$
in $T \to \infty$ and thus the
following scaled limit trajectory exists,
\[
\{ \widehat{v}(t) : t \in [0, 2] \}
= \lim_{T \to \infty}
\left\{ 
\frac{ \widehat{x}^{\rm max}_{2T}(sT)}{T} : 0 \leq s \leq 2
\right\}.
\]
This limit trajectory
exhibits a nontrivial curve.
Due to (\ref{eqn:symmetry2}), it has the following
symmetry,
\[
\widehat{v}(t)=\widehat{v} (2-t),
\quad t \in [0, 2].
\]

\subsection{Elliptic excursion processes} 
\label{sec:numerical_elli}

For the elliptic excursion process
$(\{X(t)\}_{t \in \{0,1,\dots, 2T\}}, \P^{0,0}_{2T})$,
here we consider the same parameterization
(\ref{eqn:special2}) as in the previous trigonometric process.
The behavior of the elementary weight $q(t,x)$ on $\Lambda_{2T}^{\ast}$
is similar to $\widehat{q}(t,x)$ shown by Fig. \ref{fig:tri_q}
with slight modification depending on $\kappa$.
It is a plain function of $(t.x)$ in $\Lambda_{2T}^{\ast}$.
In this special parameterization, (\ref{eqn:P1_ell_1})
and (\ref{eqn:P1_ell_c}) are written as follows,
\begin{align}
& \P^{0,0}_{2T}(X(t)=x) 
=\P^{0,0}_{2T}(X(t)=x; r, \alpha_0, \beta_0, \kappa) \Big|_{\pi r= 6T, \alpha_0/r=\pi/4, \beta_0/r=-\pi/4}
\nonumber\\
& \quad 
 = \begin{cases}
c_2(t; T) 
\vartheta_1 \left( \frac{x}{6T}; i \kappa \right)
& \\
\displaystyle{\quad \times 
\prod_{n=1}^{(t+x)/2} 
\frac{\vartheta_1 \left(\frac{1}{4} -\frac{2n+3T-2t-1}{12T}; i \kappa \right)}
{\vartheta_1 \left( \frac{n}{6T}; i \kappa \right)
\vartheta_2 \left( \frac{n}{6T}; i \kappa \right)}
\prod_{n=1}^{(t-x)/2} 
\frac{\vartheta_1 \left(\frac{1}{4}+\frac{2n+3T-2t-1}{12T}; i \kappa \right)}
{\vartheta_1 \left( \frac{n}{6T}; i \kappa \right)
\vartheta_2 \left( \frac{n}{6T}; i \kappa \right)}
} & \\
\displaystyle{
\quad \times 
\prod_{n=1}^{\{(2T-t)-x\}/2} 
\frac{\vartheta_1 \left(\frac{1}{4}+\frac{2n+2t-T-1}{12T}; i \kappa \right)}
{\vartheta_1 \left( \frac{n}{6T}; i \kappa \right)
\vartheta_2 \left( \frac{n}{6T}; i \kappa \right)}
\prod_{n=1}^{\{(2T-t)+x\}/2} 
\frac{\vartheta_1 \left(\frac{1}{4}-\frac{2n+2t-T-1}{12T}; i \kappa \right)}
{\vartheta_1 \left( \frac{n}{6T}; i \kappa \right)
\vartheta_1 \left( \frac{n}{6T}; i \kappa \right)},
} & \\
\hskip 8cm \mbox{if $(t, x) \in \Lambda_{2T}$},
& \\
0, \hskip 7.6cm \mbox{otherwise},
\end{cases}
\label{eqn:P1_ell_2}
\end{align}
with
\begin{align}
c_2(t; T)
={}&
\frac{
\prod_{n=1}^t \vartheta_1 \left( \frac{n}{6T} ; i \kappa \right) 
\prod_{n=1}^{2T-t} \vartheta_1 \left( \frac{n}{6T} ; i \kappa \right) 
}
{
\vartheta_2 \left( 0 ; i \kappa \right)
\prod_{n=1}^{2T} \vartheta_1 \left( \frac{n}{6T} ; i \kappa \right)
}
\nonumber\\
&\times
\prod_{n=1}^T
\frac{
\left\{ \vartheta_1 \left( \frac{n}{6T} ; i \kappa \right) 
\vartheta_2 \left( \frac{n}{6T} ; i \kappa \right) \right\}^2
}
{
\vartheta_1\left(\frac{1}{4}+ \frac{2n-T-1}{12T} ; i \kappa \right)
\vartheta_1\left(\frac{1}{4}- \frac{2n-T-1}{12T} ; i \kappa \right)
},
\label{eqn:P1_ell_c2}
\end{align}
where
\[
\vartheta_2(v; \tau)
\equiv \vartheta_1(v+1/2; \tau).
\]

\begin{figure}
\begin{center}
\includegraphics[width=0.5\textwidth]{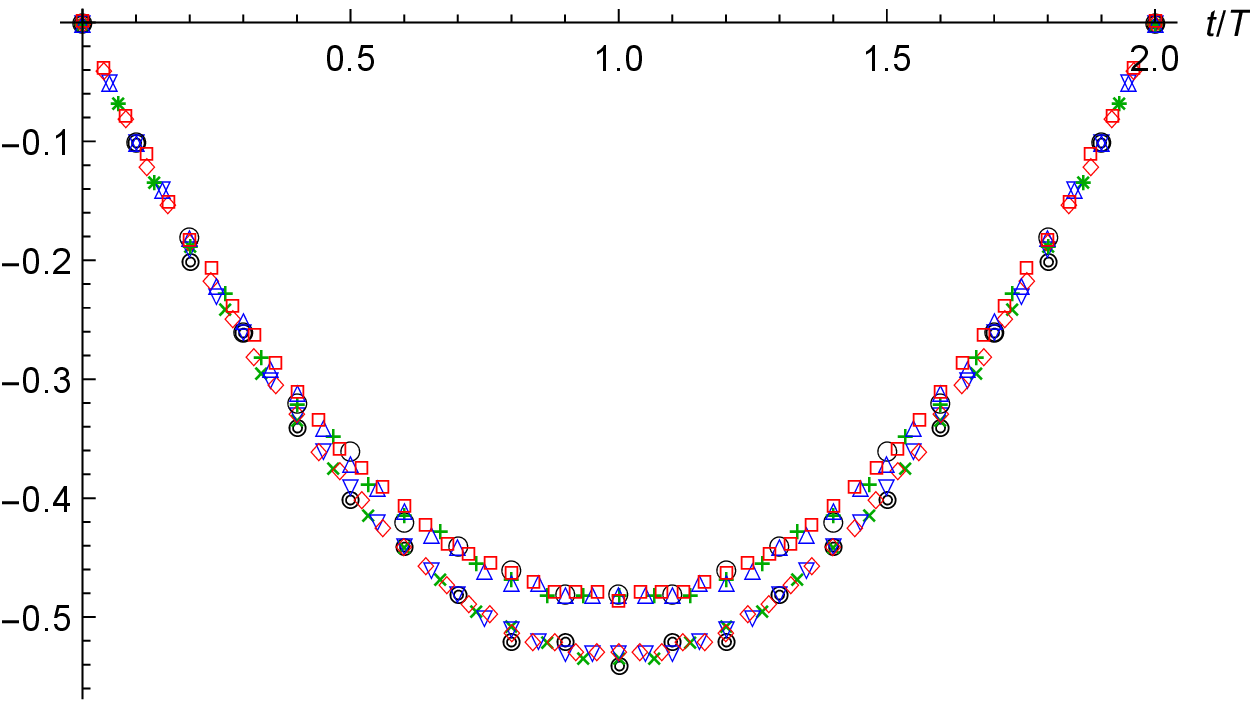}
\end{center}
\caption{
Numerical values of $x^{\rm max}_{2T}(t; \kappa)/T$, 
$t \in \{0,1, \dots, 2T\}$ are plotted for various time durations; 
for $\kappa=10$, 
$T=100$ (circles $\bigcirc$), 150 (crosses $+$), 
200 (triangles $\bigtriangleup$), 250 (squares $\Box$);
and for $\kappa=0.5$, 
$T=100$ (double circles $\circledcirc$), 150 (crosses $\times$), 
200 (reverse triangles $\bigtriangledown$), 250 (diamonds $\diamondsuit$).
The plots for $\kappa=10$ are very close to
those for the trigonometric excursion processes
shown in Figure \ref{fig:tri_max}.
The observed data-collapse suggests the existence of a limit
curve $v(t, \kappa), t \in [0, 2]$ 
of the elliptic excursion process 
in the long-term limit $T \to \infty$
for each value of $\kappa >0$.
The deviation of this limit trajectory $v(t, \kappa), t \in [0, 2]$
from the straight line is enhanced as $\kappa \to 0+$.
}
\label{fig:ell_max}
\end{figure}

When $\kappa >0$ is large, the behavior
of the elliptic excursion process
$(\{X(t)\}_{t \in \{0,1,\dots, 2T\}}, \P^{0,0}_{2T})$
is very similar to that of the trigonometric excursion
process $(\{X(t)\}_{t \in \{0,1,\dots, 2T\}}, \widehat{\P}^{0,0}_{2T})$.
Let $x^{\rm max}_{2T}(t; \kappa)$ denote
the maximum likelihood trajectory of the 
elliptic excursion process with parameter $\kappa >0$.
Figure \ref{fig:ell_max} shows
$\{x^{\rm max}_{2T}(sT; \kappa)/T : 0 \leq s \leq 2 \}$
with $\kappa=0.5$ and $\kappa=10$
for $T=100, 150, 200, 250$.
The data-collapse suggests that the values of
$x^{\rm max}_{2T}(t; \kappa)$ are proportional to $T$
in $T \to \infty$ and thus the scaled limit trajectory
\[
\{ v(s; \kappa) : s \in [0, 2] \}
= \lim_{T \to \infty}
\left\{ 
\frac{ x^{\rm max}_{2T}(sT; \kappa)}{T} : 0 \leq s \leq 2
\right\},
\]
exists for each value of $\kappa >0$. 
The deviation of this limit trajectory $v(t, \kappa), t \in [0, 2]$
from the straight line is enhanced as $\kappa \to 0+$.

\SSC
{Asymptotic Analysis of Simplified Trigonometric
Excursion Processes} \label{sec:asymptotic}

In this section, we report our trials
to analyze asymptotics of the probability law $\widetilde{\P}^{0,0}_{2T}$
of the simplified trigonometric excursion processes
in the long-term limit $T \to \infty$.

\subsection{Large deviation principle and scaled limit trajectory}
\label{eqn:LDP}

We scale the spatio-temporal coordinate
$(t, x) \in \Lambda_{2T}$ in a unit of $T$ as
\[
t=sT, \quad x= vT.
\]
Then we expect that there is a non-negative function
$I(s,v)$ such that the following asymptotics holds,
\begin{equation}
\widetilde{\P}^{0,0}_{2T}(X(sT)=vT) 
\simeq e^{-T I(s,v)}
\quad \mbox{as $T \to \infty$},
\label{eqn:LDP1}
\end{equation}
for 
\begin{align*}
\overline{\Lambda}_2
=& \{ (s, v) \in \R^2 : 0 \leq s \leq 1, -s \leq v \leq s \}
\nonumber\\
&\cup \{(s, v) \in \R^2 : 1 < s \leq 2, -(2-s) \leq v \leq 2-x \}.
\end{align*}
When the asymptotic of the form (\ref{eqn:LDP1}) is valid,
we say that the {\it large deviation principle} is established,
and the function $I$ is called the {\it rate function} 
(see, for instance, \cite{Var16}).
If so in the present system, 
the scaled limit trajectory,
which was denoted by
$v=\widetilde{v}(s), s \in [0, 2]$
and numerically studied in the previous section, will be determined
as zeros of $I$,
\begin{equation}
I(s, \widetilde{v}(s))=0, \quad
s \in [0,2],
\label{eqn:rate_func_zero}
\end{equation}
since (\ref{eqn:LDP1}) with (\ref{eqn:rate_func_zero})
implies
\begin{equation}
\lim_{T \to \infty}
\widetilde{\P}^{0,0}_{2T}(X(sT)=\widetilde{v}(s) T)=1,
\quad s \in [0, 2].
\label{eqn:wp1}
\end{equation}
This is the {\it law of large numbers} in the present problem.

As shown by (\ref{eqn:P1_tri2_2}),
$\widetilde{\P}^{0,0}_{2T}(X(sT)=vT)$ consists of
several products of trigonometric functions.
For example, it contains the product
\[
B_{2T}^{(1)}(s, v)
=\prod_{n=1}^{(s+v)T/2}
\cos \left( \frac{\{2n+(3-2s)T-1\}\pi}{6T} \right).
\]
If we take the logarithm of this, we find that
\begin{align*}
\frac{\pi}{3T} \log B_{2T}^{(1)}(s, v)
&= \frac{\pi}{3T} \sum_{n=1}^{(s+v)T/2}
\log \cos \left( \frac{\{2n+(3-2s)T-1\}\pi}{6T} \right)
\nonumber\\
&\rightarrow 
\int_{(3-2s)\pi/6}^{(3-s+v)\pi/6} \log \cos u \, du
\quad
\mbox{in $T \to \infty$}.
\end{align*}
The following Fourier expansion formulas 
are useful to evaluate the integrals of
logarithms of trigonometric functions,
\begin{align}
\int dx \, \log \sin x
&= -x \log 2 - \frac{1}{2} \sum_{n=1}^{\infty} \frac{\sin (2nx)}{n^2},
\quad 0 \leq x < \pi,
\label{eqn:Fourier1}
\\
\int dx \, \log \cos x
&= - x \log 2 + \frac{1}{2} \sum_{n=1}^{\infty} (-1)^{n-1}
\frac{\sin(2 n x)}{n^2},
\quad |x| < \frac{\pi}{2}.
\label{eqn:Fourier2}
\end{align}
We can derive the following integral representation
for the rate function.
The proof is given in Appendix \ref{sec:appendixA}. 

\begin{lem}
\label{thm:rate_func}
For the simplified trigonometric excursion process
$(\{X(t)\}_{t \in \{0,1, \dots, 2T\}}, \widetilde{\P}^{0,0}_{2T})$
with parameters {\rm (\ref{eqn:special1})}, 
the large deviation principle {\rm (\ref{eqn:LDP1})} is established
with the rate function $I$ expressed by
\begin{align}
I(s, v) =&
3 \int_0^{(1-s-v)/6} \log \frac{\sin ( \pi (\varphi+1/6 ) )}{\sin ( \pi (-\varphi+1/6 ) )} d \varphi
+ 6 \int_0^{(1-s+v)/6} \log \frac{\sin(\pi(\varphi+1/6))}{\sin ( \pi (-\varphi+1/6 ) )} d \varphi
\nonumber\\
&+ 6 \int_0^{(1-s)/3} \log \frac{\cos(\pi(\varphi+1/6))}{\cos(\pi(-\varphi+1/6))} d \varphi
\qquad 
\mbox{for $(x, v) \in \overline{\Lambda}_2$}.
\label{eqn:I1}
\end{align}
\end{lem}

By differentiating the equation (\ref{eqn:rate_func_zero})
by $s$, we obtain $dI(s, \widetilde{v}(s))/ds=0$ and then
\[
\frac{d \widetilde{v}(s)}{ds}
= - \frac{\partial_s I(s, \widetilde{v}(s))}
{\partial_v I(s, \widetilde{v}(s))},
\]
where
\[
\partial_s I(s, v)=\frac{\partial I(s, v)}{\partial s},
\quad
\partial_v I(s, v) = \frac{\partial I(s, v)}{\partial v}.
\]
Lemma \ref{thm:rate_func} gives the following
differential equation,
\[
\frac{d \widetilde{v}(s)}{d s}
= - \frac{
\displaystyle{ \log
\left[\frac{\sin(\pi(s+\widetilde{v}(s))/6) \sin^2(\pi(s-\widetilde{v}(s))/6) \cos^4 (\pi(2s-1)/6)}
{\sin(\pi(2-s-\widetilde{v}(s))/6) \sin^2 (\pi(2-s+\widetilde{v}(s))/6) \cos^4 (\pi(3-2s)/6)}
\right]
}}
{\displaystyle{ \log
\left[ \frac{\sin (\pi(s+\widetilde{v}(s))/6) \sin^2 (\pi(2-s+\widetilde{v}(s))/6)}
{\sin^2(\pi(s-\widetilde{v}(s))/6) \sin(\pi(2-s-\widetilde{v}(s))/6)}
\right]
}}.
\]

We can prove the following.
\begin{thm}
\label{thm:trajectory}
Consider the simplified trigonometric excursion process
with parameters {\rm (\ref{eqn:special1})}. 
The scaled limit trajectory 
$v=\widetilde{v}(s), s \in [0, 2]$
passes the origin at $s=1$, that is, $\widetilde{v}(1)=0$,
and in the time region in the vicinity of $s=1$,
it behaves as 
\begin{equation}
\widetilde{v}(s) = \frac{1}{3}(s-1)
-\frac{2^4 \pi^2}{3^6} (s-1)^3
+{\rm O}((s-1)^5).
\label{eqn:v_approx1}
\end{equation}
\end{thm}
\noindent{\it Proof.} \,
Assume that $|s-1| \ll 1$ and $|v| \ll 1$
in (\ref{eqn:I1}).
For $|\varphi| \ll 1$,
\begin{align*}
\sin(\pi(\varphi+1/6))
&= \frac{1}{2} \left(
1+\sqrt{3} \pi \varphi - \frac{\pi^2}{2} \varphi^2
-\frac{\sqrt{3} \pi^3}{6} \varphi^3 
+ \frac{\pi^4}{24} \varphi^4 + \frac{\sqrt{3} \pi^5}{120} \varphi^5
\right)
+{\rm O}(\varphi^6),
\nonumber\\
\cos(\pi(\varphi+1/6))
&= \frac{\sqrt{3}}{2} \left(
1- \frac{\sqrt{3} \pi}{3} \varphi - \frac{\pi^2}{2} \varphi^2
+\frac{\sqrt{3} \pi^3}{18} \varphi^3 
+ \frac{\pi^4}{24} \varphi^4 - \frac{\sqrt{3} \pi^5}{360} \varphi^5
\right)
+{\rm O}(\varphi^6).
\end{align*}
Then
\begin{align*}
& \int_0^{(1-s \pm v)/6} \log \frac{\sin ( \pi (\varphi+1/6 ) )}{\sin ( \pi (-\varphi+1/6 ) )} d \varphi
\nonumber\\
& \quad 
=\int_0^{(1-s \pm v)/6} \log \sin ( \pi (\varphi+1/6 ) ) d \varphi
- \int_{-(1-s \pm v)/6}^0 \log \sin(\pi(\varphi+1/6)) d \varphi
\nonumber\\
& \quad
= \frac{\sqrt{3} \pi}{2^2 \times 3^2}(1-s \pm v)^2 
+\frac{2 \sqrt{3} \pi^3}{2^4 \times 3^5} (1-s \pm v)^4
\nonumber\\
& \quad \quad
+ \frac{11 \sqrt{3} \pi^5}{2^4 \times 3^8 \times 5} 
(1-s \pm v)^6
+{\rm O}(|1-s \pm v|^8),
\end{align*}
and
\begin{align*}
& \int_0^{(1-s)/3} \log \frac{\cos(\pi (\varphi+1/6))}{\cos(\pi (-\varphi+1/6 ))} d \varphi
\nonumber\\
& \quad 
= \int_0^{(1-s)/3} \log \cos ( \pi (\varphi+1/6 ) ) d \varphi
- \int_{-(1-s)/3}^0 \log \cos(\pi(\varphi+1/6)) d \varphi
\nonumber\\
& \quad
= - \frac{\sqrt{3} \pi}{3^3}(1-s)^2 
-\frac{2 \sqrt{3} \pi^3}{3^7} (1-s)^4
- \frac{2^2 \sqrt{3} \pi^5}{3^9 \times 5} (1-s)^6
+{\rm O}(|1-s|^8).
\end{align*}
Together these give
\begin{align}
I(s, v) ={}&
\frac{\sqrt{3} \pi}{2^2 \times 3^2} (s-1-3v)^2
\nonumber\\
&- \frac{\sqrt{3} \pi^3}{2^3 \times 3^6}
\Big\{ 5(s-1)^4+36 v(s-1)^3
-162 v^2 (s-1)^2 + 36 v^3 (s-1) - 27 v^4 \Big\}
\nonumber\\
&- \frac{\sqrt{3} \pi^5}{2^4 \times 3^8 \times 5}
\Big\{ 29 (s-1)^6 + 198 v (s-1)^5 - 1485 v^2(s-1)^4 
\nonumber\\
& \qquad 
+660 v^3 (s-1)^3 - 1485 v^4 (s-1)^2 + 198 v^5 (s-1) -99 v^6 \Big\}
\nonumber\\
&+ {\rm O}(|s-1|^8, |v|^8).
\label{eqn:I_approx1}
\end{align}
The first line of (\ref{eqn:I_approx1}) vanishes when
$v=(s-1)/3$.
So we can assume that
\[
v=\frac{1}{3}(s-1)+ c (s-1)^3 + {\rm O}((s-1)^5)
\]
with an unknown coefficient $c$.
In this assumption (\ref{eqn:I_approx1}) is written as
\[
I(s, v)= \frac{\sqrt{3} \pi}{2^2} \left(
c+\frac{2^4 \pi^2}{3^6} \right)^2 (s-1)^6 + {\rm O}((s-1)^8), 
\]
and hence it is verified that
$I(s, \widetilde{v}(s))={\rm O}((s-1)^8)$
with (\ref{eqn:v_approx1}). 
The proof is complete.
\qed
\vskip 0.3cm

\begin{figure}
\begin{center}
\includegraphics[width=0.5\textwidth]{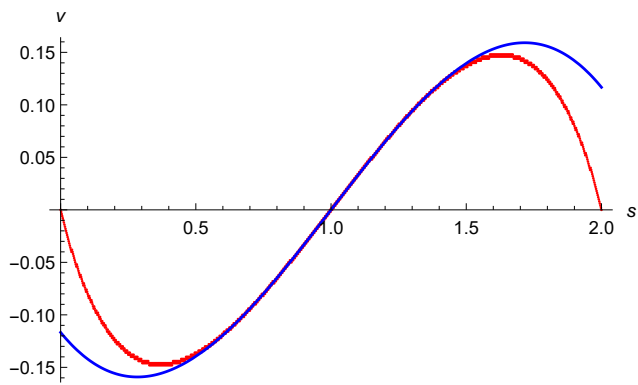}
\end{center}
\caption{
Numerical values of
$\widetilde{x}^{\rm max}_{2T}(t)/T, t \in \{0,1,\dots, 2T\}$
are plotted versus $s=t/T$ for $T=500$ by dots.
In the vicinity of $s=1$, the dots are well approximated
by the cubic curve (\ref{eqn:cubic1})
starting from $(0, -v_0)$ and ending at $(2, v_0)$
with $v_0=1/3-2^4 \pi^2/3^6 =0.116 \cdots$.
}
\label{fig:vs_curve}
\end{figure}

In Fig. \ref{fig:vs_curve}, the numerical values of
$\widetilde{x}^{\rm max}_{2T}(t)/T, t \in \{0,1,\dots, 2T\}$
with $T=500$ are plotted versus $s=t/T$
by dots, which will well simulate the
scaled limit trajectory $v=\widetilde{v}(s), s \in [0, 2]$,
since $T=500$ is large enough. 
The cubic curve
\begin{equation}
v= \frac{1}{3}(s-1)
-\frac{2^4 \pi^2}{3^6} (s-1)^3,
\quad s \in [0, 2], 
\label{eqn:cubic1}
\end{equation}
is also shown, 
which starts from $(0, -v_0)$ and ends at $(2, v_0)$
with $v_0=1/3-2^4 \pi^2/3^6 =0.116 \cdots$. 
Coincidence of this cubic curve with the dots
is excellent in the vicinity of $s=1$.

\subsection{Central limit theorem at time $t=T$}
\label{eqn:}

In Theorem \ref{thm:trajectory},
$\widetilde{v}(1)=0$ was proved.
As shown by (\ref{eqn:wp1}), 
this means that 
the trajectory with time duration $2T$ 
passes through the origin 0
at time $t=T$ with probability one, when $T \to \infty$.

Now we study the fluctuation of the trajectory
around 0 at time $t=T$.
We can prove the following {\it central limit theorem}.

\begin{prop}
\label{thm:CLT}
For the simplified trigonometric excursion process
$(\{X(t)\}_{t \in \{0,1, \dots, 2T\}}, \widetilde{\P}^{0,0}_{2T})$
with parameters {\rm (\ref{eqn:special1})}, 
\begin{equation}
\lim_{T \to \infty} \frac{\sqrt{T}}{2}
\widetilde{P}^{0,0}_{2T}(X(T)=\sqrt{T} \xi)
=f(\xi)
\label{eqn:CLT1}
\end{equation}
with
\begin{equation}
f(\xi)=\frac{3^{1/4}}{2} \exp \left( - \frac{\sqrt{3} \pi}{4} \xi^2 \right),
\quad \xi \in \R. 
\label{eqn:CLT2}
\end{equation}
That is, at time $t=T$,
the fluctuation is proportional to $\sqrt{T}$
in the limit $T \to \infty$, 
and its coefficient is normally distributed with
variance
$\sigma^2=2/(\sqrt{3} \pi)$.
\end{prop}
The proof is given in Appendix \ref{sec:appendixB}. 

Figure \ref{fig:CLT1} shows the numerical plots of 
$(\sqrt{T}/2) \widetilde{\P}^{0,0}_{2T}(X(T)=\sqrt{T} \xi)$
with parameters (\ref{eqn:special1})
for $T=100$ and the function $f(\xi)$ given by (\ref{eqn:CLT2}). 
We found that $T=100$ is large enough to see
a good coincidence. 

This result should be compared with (\ref{eqn:BM1}) 
which shows the convergence of 
the excursion process of classical random walk
to the Brownian bridge. 
We note that
the variance $\sigma^2 = 2/(\sqrt{3} \pi) =0.367 \cdots$
of the present process is smaller than
the classical value $\sigma_{\rm cl}=1/2$ at time $s=1$. 

\begin{figure}
\begin{center}
\includegraphics[width=0.5\textwidth]{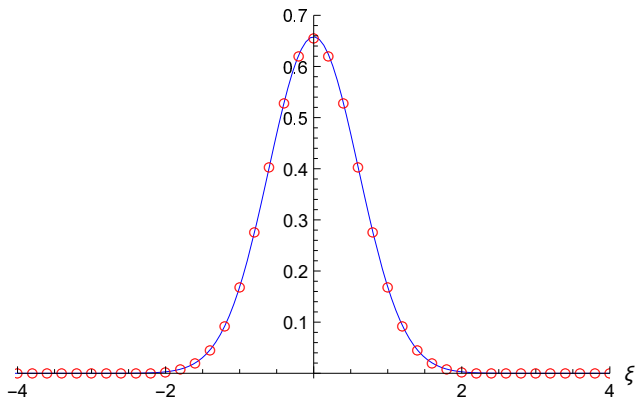}
\end{center}
\caption{
Numerical values of 
$(\sqrt{T}/2) \widetilde{\P}^{0,0}_{2T}(X(T)=\sqrt{T} \xi)$
with parameters (\ref{eqn:special1})
are plotted by circles for $T=100$ 
and the probability density function $f(\xi)$ 
of normal distribution given by (\ref{eqn:CLT2})
is drawn by a curve.
}
\label{fig:CLT1}
\end{figure}

\SSC
{Concluding Remarks} \label{sec:concluding}

The weight functions of Schlosser's lattice model \cite{Sch07}
are complex functions which are totally elliptic 
in the sense that all variables representing 
the coordinates $t, x$ and parameters $\alpha, \beta$
have equal periods of doubly periodicity,
if they are view as complex variables \cite{Spi02}.
The totally elliptic function giving the weight for the
elementary step of lattice paths (\ref{eqn:q_1}) was cleverly chosen \cite{Sch07}
so that the generating functions of lattice paths,
which are defined as summations of products of
elementary weight-functions, are completely factorized as (\ref{eqn:Q_2}).

In the present paper, we have constructed a family of one-dimensional
excursion processes from Schlosser's two-dimensional lattice path models.
Here the spatio-temporal coordinates $(t, x)$ as well as parameters are
real variables and the
weight functions are real functions of them.
In order to obtain probability measures for trajectories
of stochastic processes, we have considered 
the excursion processes and given sufficient conditions
to make the measures be non-negative definite.

Although our probability measures have lost the totally elliptic
property as complex functions,
the product formulas originally given for the generating functions
in Schlosser's lattice path models remain
in the probability measures for the excursion processes.
The obtained processes are inhomogeneous both in
space and time, and in order to describe the behavior of
trajectories, we have to precisely evaluate the ratios of
products of Jacobi's theta functions for the
elliptic excursion processes 
and of the trigonometric functions for the
reduced processes.

We have studied the excursion processes starting from the
origin and returning to the origin in time duration
$2T, T \in \N$.
If we considered the excursion process of the simple
and symmetric random walk, or its diffusion scaling limit
realized as the Brownian bridge,
the maximum likelihood trajectories are
the straight lines connecting $(0,0)$ and $(0, 2T)$
on the spatio-temporal plane.
Contrary to such classical results,
the maximum likelihood trajectories in the present 
elliptic and trigonometric excursion processes
exhibit nontrivial curves connecting $(0,0)$ and $(0, 2T)$
in the spatio-temporal plane.
Moreover, data-collapse observed in the scaling plots
for variety of time durations suggests 
the existence of scaled limit trajectories in the $T \to \infty$ limit.
There we have concentrated on the single-time measures (\ref{eqn:P1_0})
for the excursion processes. 
A general formula was given for any multi-time joint measure
by (\ref{eqn:measure1}), which provides
a well-defined probability measure for trajectories,
if the conditions (\ref{eqn:condB1})-(\ref{eqn:condB4}) of 
Theorem \ref{thm:positivity_main} are satisfied for the 
parameterization (\ref{eqn:para2}).
Spatio-temporal correlations of each trajectory should be
systematically studied in the future.
In Section \ref{sec:trajectories}, 
we reported about our numerical study only in the special cases
with $\sigma=3$ for the simplified trigonometric excursion process
and with $\sigma=6$ for the trigonometric and elliptic excursion processes,
since in these cases the degree of deviation from the classical processes $\lambda$
is maximized. Numerical study of other cases with
lower values of $\lambda$ is now in progress.

In Section \ref{sec:asymptotic}, we have reported about the asymptotic analysis
in the long-term limit $T \to \infty$ only for the 
simplified trigonometric excursion process.
There the Fourier analysis for the integrals of
logarithms of trigonometric functions are used.
In order to analyze asymptotic probability laws for the
elliptic processes, we need to develop the present method
to the elliptic level. It will be an interesting future problem.
Emergence of nontrivial curves of trajectories
shown by Figs. \ref{fig:simp_tri_T=100} and \ref{fig:tri_T=100}
from the simple elementary weight-functions 
(see Figs. \ref{fig:simp_tri_q} and \ref{fig:tri_q})
is a new aspect of the elliptic combinatorics \cite{Sch07,Sch11,SY17}
and expected to lead a way to {\it elliptic probability theory}.

In the paper by Schlosser \cite{Sch07}, 
factorization formulas are given for the
{\it Karlin-McGregor determinants} for plural 
lattice paths with the totally elliptic weights.
They are systems of {\it nonintersecting lattice paths}
weighted by complex functions.
It will be a challenging future problem
to define probability measures for 
ensembles of nonintersecting lattice paths
and construct {\it noncolliding particle systems} 
which are indeed inhomogeneous both in space and time.

\vskip 0.5cm
\noindent{\bf Acknowledgements} \quad
The present work was started
during the participation of one of the authors (MK)
in the ESI workshop on 
``Elliptic Hypergeometric Functions in Combinatorics,
Integrable Systems and Physics" (March 20-24, 2017).
MK expresses his gratitude for the hospitality of 
Erwin Schr\"odinger International Institute for Mathematics and Physics (ESI) 
of the University of Vienna
and for well-organization of the workshop by 
Christian Krattenthaler, 
Masatoshi Noumi, 
Simon Ruijsenaars, 
Michael J. Schlosser, 
Vyacheslav P. Spiridonov, 
and S. Ole Warnaar.
The authors thank Michael J. Schlosser for useful discussion.
This work was supported in part by
the Grant-in-Aid for Scientific Research (C) (No.26400405),
(B) (No.26287019), 
(B) (No.18H01124), 
and
(S) (No.16H06338) of Japan Society for
the Promotion of Science.

\vskip 1cm
\appendix
\SSC{Proof of Lemma \ref{thm:rate_func}}
\label{sec:appendixA}
Let
\begin{align*}
&A_{2T}(\gamma) = \prod_{n=1}^{\gamma T} \sin \left( \frac{n \pi}{3T} \right),
\nonumber\\
&B_{2T}^{(1)}(s, v) = \prod_{n=1}^{(s+v)T/2}
\cos \left( \frac{\{2n+(3-2s)T-1\}\pi}{6T} \right),
\nonumber\\
&B_{2T}^{(2)}(s, v) = \prod_{n=1}^{\{(2-s)-v\}T/2}
\cos \left( \frac{\{2n+(2s-1)T-1\}\pi}{6T} \right),
\nonumber\\
&B_{2T}^{(3)} = \prod_{n=1}^{T}
\cos \left( \frac{(2n-T-1)\pi}{6T} \right).
\end{align*}
For fixed $\gamma < \infty$, 
\begin{align*}
\frac{\pi}{3T} \log A_{2T}(\gamma)
&= \frac{\pi}{3T} \sum_{n=1}^{\gamma T} 
\log \sin \left( \frac{n \pi}{3T} \right)
\, \longrightarrow \, 
\int_0^{\gamma \pi/3} \log \sin u \, dy
\quad \mbox{in $T \to \infty$}.
\end{align*}
If we use (\ref{eqn:Fourier1}), we have the
following expression for the limit,
\[
- \lim_{T \to \infty} \frac{1}{T} \log A_{2T}(\gamma)
=\gamma \log 2 
+ \frac{3}{2 \pi} \sum_{n=1}^{\infty} \frac{1}{n^2}
\sin \left( \frac{2 \gamma n \pi}{3} \right).
\]
Similarly, we obtain the following limits,
\begin{align*}
& - \lim_{T \to \infty} \frac{1}{T}
\log B_{2T}^{(1)}(s, v)
\nonumber\\
& \qquad 
= \frac{s+v}{2} \log 2
- \frac{3}{2 \pi} \sum_{n=1}^{\infty}
\frac{(-1)^{n-1}}{n^2}
\left[ \sin \left( \frac{(3-s+v)n \pi}{3} \right)
- \sin \left( \frac{(3-2s)n \pi}{3} \right) \right],
\nonumber\\
& - \lim_{T \to \infty} \frac{1}{T}
\log B_{2T}^{(2)}(s, v)
\nonumber\\
& \qquad 
= \frac{2-s-v}{2} \log 2
- \frac{3}{2 \pi} \sum_{n=1}^{\infty}
\frac{(-1)^{n-1}}{n^2}
\left[ \sin \left( \frac{(1+s-v)n \pi}{3} \right)
- \sin \left( \frac{(2s-1)n \pi}{3} \right) \right],
\nonumber\\
& - \lim_{T \to \infty} \frac{1}{T}
\log B_{2T}^{(3)}
= \log 2
- \frac{3}{\pi} \sum_{n=1}^{\infty}
\frac{(-1)^{n-1}}{n^2}
\sin \left( \frac{n \pi}{3} \right). 
\end{align*}
After some calculation using the addition formulas
of trigonometric functions and the equality
\begin{equation}
\sum_{n=1}^{\infty} \frac{(-1)^{n-1}}{n^2} 
\sin \left( \frac{n \pi}{3} \right)
=\frac{2}{3} \sum_{n=1}^{\infty} \frac{1}{n^2}
\sin \left( \frac{n \pi}{3} \right), 
\label{eqn:sin_equality}
\end{equation}
we arrive at the
following Fourier expansion for the
rate function,
\begin{align}
I(s,v) \equiv& 
-\lim_{T \to \infty} \frac{1}{T}
\log \widetilde{\P}^{0,0}_{2T}(X(st)=vT)
\nonumber\\
={}& \frac{6}{\pi} \sum_{n=1}^{\infty} \frac{(-1)^{n-1}}{n^2}
\sin \left( \frac{n \pi}{3} \right) \cos \left( \frac{2(1-s)n \pi}{3} \right)
+ \frac{5}{\pi} \sum_{n=1}^{\infty} \frac{1}{n^2} \sin \left( \frac{n \pi}{3} \right)
\nonumber\\
&- \frac{6}{\pi} \sum_{n=1}^{\infty} \frac{1}{n^2}
\sin \left( \frac{n \pi}{3} \right) \cos \left( \frac{(1-s+v)n \pi}{3} \right)
\nonumber\\
&- \frac{3}{\pi} \sum_{n=1}^{\infty} \frac{1}{n^2}
\sin \left( \frac{n \pi}{3} \right) \cos \left( \frac{(1-s-v)n \pi}{3} \right).
\label{eqn:I_Fourier}
\end{align}
Let
\[
J(\varphi) = \sum_{n=1}^{\infty} \frac{1}{n^2} 
\sin \left( \frac{n \pi}{3} \right) \cos (n \varphi),
\quad
K(\varphi) = \sum_{n=1}^{\infty} \frac{(-1)^{n-1}}{n^2} 
\sin \left( \frac{n \pi}{3} \right) \cos (n \varphi).
\]
Then by (\ref{eqn:Fourier1}) and (\ref{eqn:Fourier2}), we have
\begin{align*}
J(\varphi_1, \varphi_2) &\equiv J(\varphi_1)-J(\varphi_2)
\nonumber\\
&= -\frac{1}{2} \left[
\int_{\pi/3+\varphi_2}^{\pi/3+\varphi_1}
\log \sin \frac{u}{2} \, du
- \int_{\pi/3-\varphi_1}^{\pi/3-\varphi_2}
\log \sin \frac{u}{2} \, du \right]
\nonumber\\
&= -\frac{1}{2} \left[
\int_{\varphi_2}^{\varphi_1}
\log \sin \left( \frac{u}{2} + \frac{\pi}{6} \right) \, du
- \int_{-\varphi_1}^{-\varphi_2}
\log \sin \left( \frac{u}{2} + \frac{\pi}{6} \right) \, du \right],
\end{align*}
\begin{align*}
K(\varphi_1, \varphi_2) &\equiv K(\varphi_1)-K(\varphi_2)
\nonumber\\
&= \frac{1}{2} \left[
\int_{\pi/3+\varphi_2}^{\pi/3+\varphi_1}
\log \cos \frac{u}{2} \, du
- \int_{\pi/3-\varphi_1}^{\pi/3-\varphi_2}
\log \cos \frac{u}{2} \, du \right]
\nonumber\\
&= \frac{1}{2} \left[
\int_{\varphi_2}^{\varphi_1}
\log \cos \left( \frac{u}{2} + \frac{\pi}{6} \right) \, du
- \int_{-\varphi_1}^{-\varphi_2}
\log \cos \left( \frac{u}{2} + \frac{\pi}{6} \right) \, du \right].
\end{align*}
Note that the equality (\ref{eqn:sin_equality}) is written as
$K(0)=(2/3) J(0)$.
Then we can show that (\ref{eqn:I_Fourier}) is rewritten as
\begin{align}
I(s,v) =& - \frac{3}{\pi} J \left( \frac{(1-s-v)\pi}{3}, \frac{(1-s)\pi}{3} \right)
\nonumber\\
&- \frac{5}{\pi} J \left( \frac{(1-s+v)\pi}{3}, 0 \right)
+ \frac{1}{\pi} J \left( \frac{(1-s)\pi}{3}, \frac{(1-s+v)\pi}{3}  \right)
\nonumber\\
& - \frac{4}{\pi} J \left(\frac{(1-s)\pi}{3}, 0 \right)
+ \frac{6}{\pi} K \left(\frac{2(1-s)\pi}{3}, 0 \right) 
\nonumber\\
=& - \frac{3}{\pi} J \left( \frac{(1-s-v) \pi}{3}, 0 \right)
-\frac{6}{\pi} J \left( \frac{(1-s+v) \pi}{3}, 0 \right)
\nonumber\\
&+ \frac{6}{\pi} K \left( \frac{2(1-s) \pi}{3}, 0 \right).
\label{eqn:I_JK}
\end{align}
It is easy to verify that (\ref{eqn:I_JK}) is equal to
(\ref{eqn:I1}). The proof is hence complete. \qed

\SSC{Proof of Proposition \ref{thm:CLT}}
\label{sec:appendixB}

By a similar calculation to that given in
the proof of Lemma \ref{thm:rate_func} in Appendix \ref{sec:appendixA},
we can show that
\begin{align*}
- \frac{1}{T} \log \widetilde{\P}^{0,0}_{2T}(X(T)=x)
&=
\frac{27}{2 \pi} \sum_{n=1}^{\infty} \frac{(-1)^{n-1}}{n^2} 
\sin \left( \frac{n \pi}{3} \right)
\nonumber\\
& \qquad 
-\frac{9}{\pi} \sum_{n=1}^{\infty} \frac{1}{n^2}
\sin \left( \frac{n \pi}{3} \right)
\cos \left( \frac{x n \pi}{3T} \right) 
+{\rm o}(T)
\quad \mbox{in $T \to \infty$}.
\end{align*}
Since the equality (\ref{eqn:sin_equality})
holds, the above is written as
\begin{align}
& \frac{9}{\pi} 
\sum_{n=1}^{\infty} \frac{1}{n^2}
\sin \left( \frac{n\pi}{3} \right)
\left\{ 1- \cos \left( \frac{x n \pi}{3T} \right) \right\} 
+{\rm o}(T)
\nonumber\\
& \quad =\frac{9}{\pi} \left[
\sum_{n=1}^{\infty} \frac{1}{n^2}
\sin \left( \frac{n\pi}{3} \right)
\right.
\nonumber\\
& \quad \qquad \left.
- \frac{1}{2} \sum_{n=1}^{\infty} \frac{1}{n^2}
\sin \left( \frac{n \pi}{3} 
\left(1+ \frac{x}{T} \right) \right)
- \frac{1}{2} \sum_{n=1}^{\infty} \frac{1}{n^2}
\sin \left( \frac{n \pi}{3} 
\left(1- \frac{x}{T} \right) \right)
\right]+{\rm o}(T).
\label{eqn:eqA1}
\end{align}
Using (\ref{eqn:Fourier1}), we can verify that
(\ref{eqn:eqA1}) is equal to
\begin{align*}
& \frac{9}{2 \pi} \left[
- \int_0^{\pi/3} \log \sin \frac{u}{2} \, du
+ \frac{1}{2} \int_0^{\pi(1+x/T)/3}
\log \sin \frac{u}{2} \, du
+ \frac{1}{2} \int_0^{\pi(1-x/T)/3}
\log \sin \frac{u}{2} \, du \right]
\nonumber\\
& \quad
= \frac{9}{2 \pi} \left[
\int_0^{\pi x/3T} 
\log \sin \left( \frac{u}{2} + \frac{\pi}{6} \right) du
- \int_{-\pi x/3T}^0 
\log \sin \left( \frac{u}{2} + \frac{\pi}{6} \right) du
\right].
\end{align*}
Now we put
$x=\sqrt{T} \xi$.
Since
\[
\log \sin \left( \frac{u}{2}+ \frac{\pi}{6} \right)
= - \log 2 + \frac{\sqrt{3}}{2} u
- \frac{1}{2} u^2 + {\rm O}(u^3),
\]
we obtain the evaluation
\[
\log \widetilde{\P}^{0,0}_{2T}(X(T)=\sqrt{T} \xi)
= - \frac{\sqrt{3} \pi}{4} \xi^2 +
{\rm O} (T^{-2})
\quad \mbox{in $T\to \infty$}.
\]
This implies that
\[
\widetilde{\P}^{0,0}_{2T}(X(T)=\sqrt{T} \xi)
\simeq \mbox{const.} \times
\exp \left( - \frac{\sqrt{3} \pi}{4} \xi^2 \right)
\quad \mbox{in $T \to \infty$}.
\]
The constant factor is determined by the normalization
condition and the probability density function
(\ref{eqn:CLT2}) is obtained.
The proof is thus complete. \qed



\end{document}